\RequirePackage[]{graphicx} 

\def\mode{1} 

\documentclass[journal]{article}

\usepackage{etoolbox}
\newtoggle{MRM}
\newtoggle{ARXIV}

\newtoggle{SUPMATERIAL}
\togglefalse{SUPMATERIAL}

\if 1\mode
	\toggletrue{ARXIV}
	\togglefalse{MRM}
\fi

\if 2\mode
	\toggletrue{MRM}
	\togglefalse{ARXIV}
\fi

\iftoggle{MRM}{
	\usepackage[nomarkers, nolists]{endfloat}
}

\usepackage[utf8]{inputenc}
\usepackage{amsmath}
\usepackage{bm}
\usepackage{bbold}
\usepackage{authblk}
\usepackage{amssymb}
\usepackage{graphicx}
\usepackage{ifdraft}
\usepackage{marginnote}
\usepackage{pdfpages}
\usepackage{tabularx,booktabs}
\usepackage{rotating} 
\usepackage{placeins}
\usepackage{ifpdf}
\ifpdf
	\usepackage[]{hyperref}
\else
	\usepackage[dvipdfm]{hyperref}
\fi
\usepackage{tikz}
\usepackage[capitalise]{cleveref}

\usepackage{blindtext}

\usepackage[
per-mode=fraction,
separate-uncertainty,
retain-unity-mantissa=false,
product-units=power,
table-alignment=center,
detect-all,
binary-units=true,
exponent-product=\cdot,
exponent-base=10
]{siunitx}
\usepackage{color}

\usepackage{changes}
\newcommand{\stkout}[1]{\ifmmode\text{\sout{\ensuremath{#1}}}\else\sout{#1}\fi}
\setdeletedmarkup{\stkout{#1}}

\title{Quantitative Magnetic Resonance Imaging by Nonlinear Inversion of the Bloch Equations}

\newcommand{\authorA}{Nick Scholand}
\newcommand{\authorB}{Xiaoqing Wang}
\newcommand{\authorC}{Volkert Roeloffs}
\newcommand{\authorD}{Sebastian Rosenzweig}
\newcommand{\authorE}{Martin Uecker}

\newcommand{\affilA}{Institute of Biomedical Imaging,
	Graz University of Technology, Graz, Austria}
\newcommand{\affilB}{Institute for Diagnostic and Interventional Radiology,
	University Medical Center Göttingen, Göttingen, Germany}
\newcommand{\affilC}{German Centre for Cardiovascular Research (DZHK),
	Partner Site Göttingen, Göttingen, Germany}
\newcommand{\affilD}{Cluster of Excellence “Multiscale Bioimaging: from Molecular Machines to Networks of Excitable Cells” (MBExC),University of Göttingen, Göttingen, Germany}

\newcommand{\corAdress}{Nick Scholand,
Institute of Biomedical Imaging,
Graz University of Technology,
Stremayrgasse 16/III,
8010 Graz, Austria. }

\newcommand{\corMail}{scholand@tugraz.at}
\newcommand{\runHead}{Scholand et al.}

\iftoggle{ARXIV}{
		\author[1,3]{\authorA \thanks{\corAdress \corMail}}
	}

\iftoggle{MRM}{
		\author[1,3]{\authorA}
	}

\author[1,3]{\authorB}
\author[2]{\authorC}
\author[2,3]{\authorD}
\author[1,2,3,4]{\authorE}

\affil[1]{\affilA}
\affil[2]{\affilB}
\affil[3]{\affilC}
\affil[4]{\affilD}
	
\iftoggle{MRM}{
	\papertype{Note}
	\corraddress{\corAdress}
	\corremail{\corMail\\
		
			This work was supported by DZHK (German Centre for Cardiovascular Research),
			funded in part by DFG (German Research Foundation) under
			Germany’s Excellence Strategy - EXC 2067/1- 390729940, and
			in part by \fundingAgency{NIH} under grant \fundingNumber{U24EB029240}.\\
			
			Submitted to XXX
			\bf{{Word count}}: Abstract , Body .}
	
	\runningauthor{\runHead}
}

\begin{document}

\maketitle

\begin{center}
	\large
	Submitted to Magnetic Resonance in Medicine
\end{center}

\begin{abstract}
\noindent\textbf{Purpose:}
Development of a generic model-based reconstruction framework for
multi-parametric quantitative MRI that can be used with data
from different pulse sequences.

\noindent\textbf{Methods:}
Generic nonlinear model-based reconstruction for quantitative MRI
estimates parametric maps directly from the acquired k-space by
numerical optimization.
This requires numerically accurate and efficient methods to solve
the Bloch equations and their partial derivatives.
In this work, we combine direct sensitivity analysis and
pre-computed state-transition matrices into a generic framework
for calibrationless model-based reconstruction that can be applied
to different pulse sequences. As a proof-of-concept, the method
is implemented and validated for quantitative $T_1$ and $T_2$
mapping with single-shot inversion-recovery (IR) FLASH and
IR bSSFP sequences in simulations, phantoms, and the human brain.

\noindent\textbf{Results:} 
The direct sensitivity analysis enables a highly accurate and numerically
stable calculation of the derivatives.
The state-transition matrices efficiently exploit repeating patterns in
pulse sequences, speeding up the calculation by a factor of 10 for the
examples considered in this work,
while preserving the accuracy of native ODE solvers.
The generic model-based method reproduces quantitative results of
previous model-based reconstructions based on the known
analytical solutions for radial IR FLASH. For IR bSFFP it
produces accurate $T_1$ and $T_2$ maps for the NIST phantom
in numerical simulations and experiments. Feasibility is also
shown for human brain, although results are affected by
magnetization transfer effects.

\noindent\textbf{Conclusion:}
By developing efficient tools for numerical optimizations using
the Bloch equations as forward model, this work enables generic
model-based reconstruction for quantitative MRI.\\[2pt]

\end{abstract}


\section{Introduction}
\label{sec1}

Conventional quantitative magnetic resonance imaging is based on
a two-step process, where first intermediate images are reconstructed and
then physical models are fitted pixel-wisely to obtain parameter maps.
Acquiring a sufficient amount of high quality images with carefully designed
contrasts is required for achieving a good fit. For this reasons, these
methods are too slow for many clinical applications. In contrast, nonlinear
model-based reconstruction methods formulate image reconstruction as a
single inverse problem.
They exploit a physical model of the measurement process
and directly estimate quantitative parameter maps from k-space.
Thus, they make optimal use of the available data and enable
highly efficient parameter mapping from signals acquired with
sequences that make use of transient magnetization dynamics
\cite{Graff__2006,
	Olafsson_IEEETrans.Med.Imag._2008,
	Block_IEEETrans.Med.Imaging_2009,
	Sumpf_J.Magn.Reson.Imaging_2011,
	Wang_Philos.Trans.R.Soc.A._2021}.
These techniques have two problems: They are computationally
demanding and they need to be specially designed for each
application.

Alternatively, fingerprinting \cite{Ma_Nature_2013} uses a lookup
dictionary obtained by Bloch simulations to map the
pixels of intermediate images computed directly from undersampled data to
quantitative parameter maps. This enables multi-parametric mapping
with high acceleration in a flexible and computationally efficient
framework, but is not optimal due to its lack of a least-squares data 
consistency term.
Subspace models can be exploited for a more efficient mapping
by approximating the physical signal with a larger
linear subspace.
They reduce the computational demand of the reconstruction very
efficiently \cite{Petzschner_Magn.Reson.Med._2011,huang_Magn.Reson.Med.2012,Zhao_Magn.Reson.Med._2015,Tamir_Magn.Reson.Med._2017,Asslaender_Magn.Reson.Med._2018},
but are still
not optimal because a linear subspace is used to approximate the
manifold of possible signals.  For complicated
spin dynamics a larger number of subspace coefficients may be needed
to accurately represent the signal, rendering subspace methods
less efficient \cite{Wang_Philos.Trans.R.Soc.A._2021}.

The aim of this work is to develop a generic framework for
nonlinear model-based reconstruction with accurate signal models
for different MRI sequences even with complicated spin dynamics.
The generalization of the forward model then allows the use of 
optimized sequences for which no analytical
expression for their signal can be derived.
Fundamentally, such a generic method requires 
efficient techniques to compute the partial derivatives of
the Bloch equations.
So far, two different methods were used in MRI.
First, symbolic derivatives can be calculated for analytical solutions of the Bloch equations
for special sequences \cite{Hilbert_J.Magn.Reson.Imag._2018,Wang_Magn.Reson.Med._2018,Tan_Magn.Reson.Med._2019}
and this can be generalized to chains of small blocks using automatic differentiation \cite{Lee_Magn.Reson.Med._2019,Dong_SIAMJ.Imag.Sci._2019}.
These methods require idealized assumptions
such as hard pulse approximation and perfect inversion. In more complicated scenarios
with long pulses or imperfect inversion they require high discretization rates
or suffer from errors.
Second, derivatives can be estimates using difference quotients \cite{Ben-Eliezer_Magn.Reson.Med._2016,Sbrizzi_MagnResonImaging_2018}.
This method fully exploits the generality of full Bloch simulations,
but is computational expensive and requires careful tuning to balance
accuracy and noise amplification.

To overcome these limitations, this work uses a
direct sensitivity analysis \cite{Dickinson_J.Comput.Phys._1976}
to compute the derivatives of the Bloch equations 
for arbitrary sequences with high accuracy by solving an extended
system of ordinary differential equations (ODE). 
The technique is validated using an analytical model
of an IR bSSFP sequence and is compared to results
obtained from difference quotients methods.
To further improve computation
speed, pre-computed state-transition matrices are applied to
arbitrary initial conditions solving the required
ODEs for all repeating parts of an MRI sequence efficiently.
They are validated by comparing them to the direct application of
a Runge-Kutta ODE solver. We further integrate both techniques
in a nonlinear model-based reconstruction with integrated
calibration-less parallel imaging. For IR FLASH, we show
that the methods reproduce the results of an analytical
model. 
In a numerical and measured phantom study with an IR bSSFP
sequence we refine the flexible forward model of the generic model-based
reconstruction to include realistic simulations with slice-selective
excitations and hyperbolic secant inversion pulses.
Thus, we show that the reconstruction quality benefits
much from the more physically accurate modelling leading
to accurate $T_1$ and $T_2$ parameter maps.
Finally, we test the developed technique on in-vivo brain data
from a healthy volunteer.

Parts of this work have been published in
\cite{Scholand__2019a,Scholand__2020a,Scholand__2022a,Scholand__2022b}.

\section{Theory}

In the following, we briefly explain the concepts of a direct sensitivity
analysis and its application to the Bloch equations (SAB). We then 
describe how state-transition matrices (STMs) can be used
to accelerate the solution of the ODEs. Afterwards, both methods
are integrated into a nonlinear model-based reconstruction
method.

\subsection{Sensitivity Analysis of the Bloch Equations}
\label{sec::sa}

We consider the temporal evolution of a magnetization
vector $\boldsymbol{M}(\boldsymbol{c},t)$ depending 
on a vector of parameters $\boldsymbol{c}$ and time $t$.
The temporal behaviour of its components $M_i(\boldsymbol{c},t)$ 
is described by the Bloch equations as a system of ODEs
\begin{equation}
	\frac{\text{d}}{\text{dt}}M_i(\boldsymbol{c},t) = f_i(\boldsymbol{M}(\boldsymbol{c},t),\boldsymbol{c},t)~,
	\label{eq::bloch_ode}
\end{equation}
where $f$ defines the dynamics.
The partial derivative of the component $M_i$ 
with respect to the parameter $\boldsymbol{c}_j$
defines the $(i,j)$-th entry 
\begin{equation}
	Z_{ij}(t)=\frac{\partial{M_i(\boldsymbol{c},t)}}{\partial{{c}_j}}
\end{equation}
of the sensitivity matrix $\boldsymbol{Z}(t)$.

Using direct sensitivity analysis \cite{Dickinson_J.Comput.Phys._1976}
one obtains $\boldsymbol{Z}(t)$ by solving an additional set of ODEs.
Assuming that the partial and ordinary derivatives interchange,
the time derivative of the $(i,j)$-th entry of $\boldsymbol{Z}$ is
\begin{equation}
	\frac{\text{d}}{\text{d}t}Z_{ij}(t)
	=\frac{\text{d}}{\text{d}t}\left(\frac{\partial{M_i}(\boldsymbol{c},t)}{\partial{{c}_j}}\right)
	=\frac{\partial}{\partial{{c}_j}}\left(\frac{\text{d}{M_i}(\boldsymbol{c},t)}{\text{d}t}\right)~.
\end{equation}
Substituting $\frac{\text{d}{M_i}(\boldsymbol{c},t)}{\text{d}{t}}$ then yields
\begin{equation}
	\frac{\text{d}}{\text{d}t}Z_{ij}(t)=\frac{\partial}{\partial{\boldsymbol{c}_j}}f_i(\boldsymbol{M}(\boldsymbol{c},t),\boldsymbol{c},t)~.
\end{equation}
With the chain rule, the resulting ODE becomes
\begin{align}
	\frac{\text{d}}{\text{d}t}Z_{ij}(t)
	&=\frac{\partial{f_i}(\boldsymbol{M}(\boldsymbol{c},t),t,{\boldsymbol{c}})}{\partial{\boldsymbol{c}_j}}+\sum\limits_{j}\frac{\partial{f_i}(\boldsymbol{M}(\boldsymbol{c},t),{\boldsymbol{c}},t)}{\partial{M_j}}\frac{\partial{M_j}(\boldsymbol{c},t)}{\partial{\boldsymbol{c}_j}}\nonumber\\
	&=\frac{\partial{f_i}(\boldsymbol{M}(\boldsymbol{c},t),{\boldsymbol{c}},t)}{\partial{\boldsymbol{c}_j}}+\sum\limits_{j}\frac{\partial{f_i}(\boldsymbol{M}(\boldsymbol{c},t),{\boldsymbol{c},t})}{\partial{M_j}}Z_{ij},
\end{align}
where $\frac{\partial{f_i}(\boldsymbol{M}(\boldsymbol{c},t),{\boldsymbol{c}},t)}{\partial{M_j}}$ describes the $(i,j)$-th element of the Jacobian $J_{i,j}$. 
This can be written compactly for the sensitivity matrix $\boldsymbol{Z}(t)$ as
\begin{equation}
	\frac{\text{d}}{\text{d}t}\boldsymbol{Z}(t)=\boldsymbol{f_c}(\boldsymbol{M}(\boldsymbol{c},t),{\boldsymbol{c}},t)+\boldsymbol{J}(\boldsymbol{M}(\boldsymbol{c},t),{\boldsymbol{c}},t)\cdot\boldsymbol{Z}(t)~.
	\label{eq::sensitivity_ode}
\end{equation}
If a direct sensitivity analysis is applied to the Bloch equations for the parameters $R_1$, $R_2$ and $B_1$,
the ODE in Eqs.~\ref{eq::sensitivity_ode} describing the temporal evolution of the sensitivities becomes
\begin{equation}
	\scalebox{.6}{$
		\frac{\text{d}}{\text{d}t}\boldsymbol{Z}
		=
		\begin{pmatrix}
			0&-M_x&-\gamma\sin\phi~M_z\\
			0&-M_y&\gamma\cos\phi~M_z\\
			M_0-M_z&0&\gamma(\sin\phi~M_x-\cos\phi~M_y)
		\end{pmatrix}
		+
		\begin{pmatrix}
			-R_2&\gamma{B_z}&-\gamma\sin\phi~ B_1\\
			-\gamma{B_z}&-R_2&\gamma\cos\phi~ B_1\\
			\gamma\sin\phi B_1&-\gamma\cos\phi~ B_1&-R_1
		\end{pmatrix}
		\cdot\boldsymbol{Z}~.$}
	\label{eq::sens_ode}
\end{equation}
depending on the $x$, $y$ and $z$ components of the magnetization $\boldsymbol{M}$,
the magnetic fields $B_{z}$ and $B_1$ as well as the RF pulse phase $\phi$.
Eq.~\ref{eq::sens_ode} is solved jointly with the
Bloch Eqs.~\ref{eq::bloch_ode} which provide
the time-dependent solutions for $M_x$, $M_y$, and $M_z$.

\subsection{State-Transition Matrices}
\label{sec:stm}

By embedding the magnetization vector into a four-dimensional space
\begin{equation}
	\boldsymbol{M}(\boldsymbol{c},t)=
	\begin{pmatrix}
		M_x(\boldsymbol{c},t)\\
		M_y(\boldsymbol{c},t)\\
		M_z(\boldsymbol{c},t)\\
		1
	\end{pmatrix},
\end{equation}
we obtain a formulation of the Bloch Eqs.~\ref{eq::bloch_ode}
as a system of homogeneous ODEs
\begin{equation}
	\frac{\textrm{d}\boldsymbol{M}(\boldsymbol{c},t)}{\textrm{d}t} 
	= \boldsymbol{f}(\boldsymbol{M}(\boldsymbol{c},t),\boldsymbol{c},t)
	= \boldsymbol{A}(\boldsymbol{c},t)~\boldsymbol{M}(\boldsymbol{c},t)~,
	\label{eq::bloch_homo}
\end{equation}
with the system matrix
\begin{equation}
	\boldsymbol{A}(\boldsymbol{c},t) =
	\begin{pmatrix}
		-R_2&\gamma\boldsymbol{G}_z(t)\cdot\boldsymbol{r}&-\gamma B_y(t)&0\\
		-\gamma\boldsymbol{G}_z(t)\cdot\boldsymbol{r}&-R_2&\gamma B_x(t)&0\\
		\gamma B_y(t)&-\gamma B_x(t)&-R_1&M_0R_1\\
		0&0&0&0\\
	\end{pmatrix},
	\label{eq::Bloch_system_matrix}
\end{equation}
at location $\boldsymbol{r}$ depending on time $t$, the $z$-gradient $\boldsymbol{G}_z$ and the magnetic fields $B_{x,y}$.

The Bloch equations can be solved directly for time-dependent
coefficients $\boldsymbol{A}(t)$ using standard ODE solvers.
Here we describe the pre-computation of STMs as more efficient
way to solve the equations for MRI sequences with repeating patterns.
A STM $\boldsymbol{S}_{t_1\rightarrow t_2}$ describes the evolution 
of an arbitrary starting magnetization $\boldsymbol{M}(\boldsymbol{c},t_1)$ for
the time span from $t_1$ to $t_2$ including all effects from
relaxation and time-dependent external RF fields.
This compresses the temporal evolution to a single matrix
multiplication
\begin{equation}
	\boldsymbol{M}(\boldsymbol{c},t_2)
	=\boldsymbol{S}_{t_1\rightarrow t_2}\boldsymbol{M}(\boldsymbol{c},t_1)~.
	\label{eq::stm}
\end{equation}
The computation of $\boldsymbol{S}_{t_1\rightarrow t_2}$ is 
based on derivation of Eq.~\ref{eq::stm}
\begin{equation}
	\frac{\text{d}\boldsymbol{M}(\boldsymbol{c},t_2)}{\text{d}t_2}
	=\frac{\text{d}}{\text{d}t_2}\left(\boldsymbol{S}_{t_1\rightarrow t_2}\boldsymbol{M}(\boldsymbol{c},t_1)\right)~.
	\label{eq::stm_ode_d1}
\end{equation}
Using the Bloch Eqs.~\ref{eq::bloch_homo} to replace the
time derivative on the left side and using 
\mbox{$\frac{\text{d}\boldsymbol{M}(t_1)}{\text{d}t_2}=0$}
for the right, we obtain
\begin{equation}
	\boldsymbol{A}(t_2)\boldsymbol{M}(\boldsymbol{c},t_2)
	= \left( \frac{\text{d}}{\text{d}t_2}\boldsymbol{S}_{t_1\rightarrow t_2}\right)~\boldsymbol{M}(\boldsymbol{c},t_1)~.
	\label{eq::stm_ode_d2}
\end{equation}
By using Eqs.~\ref{eq::stm} and switching both sides we obtain
\begin{equation}
	\frac{\text{d}}{\text{d}t_2}\boldsymbol{S}_{t_1\rightarrow t_2}~\boldsymbol{M}(\boldsymbol{c},t_1)
	=\boldsymbol{A}(t_2)\boldsymbol{S}_{t_1\rightarrow t_2}\boldsymbol{M}(\boldsymbol{c},t_1)~.
	\label{eq::stm_ode_d3}
\end{equation}
As this holds for arbitrary $\boldsymbol{M}(\boldsymbol{c},t_1)$ 
and by renaming $t_2$ as $t$ a system of ODEs 
\begin{equation}
	\frac{\text{d}}{\text{d}t}\boldsymbol{S}_{t_1\rightarrow t}
	=\boldsymbol{A}(t)\boldsymbol{S}_{t_1\rightarrow t}
	\label{eq::stm_ode}
\end{equation}
for the entries of the STM is derived.
This ODEs \ref{eq::stm_ode} can be solved for estimating 
$\boldsymbol{S}_{t_1\rightarrow t_2}$ column-wisely 
with an ODE solver \cite{Moler_SIAMRev._1978,Moler_SIAMRev._2003}
with initial conditions
\begin{equation}
	\boldsymbol{S}_{t_1\rightarrow t_1}=\mathbb{1}~.
\end{equation}
The solution of the state-transition ODE in Eqs.~\ref{eq::stm_ode}
can be formally defined as a time ordered exponential
\begin{align}
	\boldsymbol{S}_{t_1 \rightarrow t_2}
	&=\prod\limits_{t_1}^{t_2}e^{\boldsymbol{A}(\tau)\text{d}\tau}
	\equiv \mathcal{T}\left\{e^{\int_{t_1}^{t_2}\boldsymbol{A}(\tau)\text{d}\tau}\right\}\\
	&\equiv \underset{N\rightarrow\infty}{\text{lim}}
	\left(e^{\boldsymbol{A}(t_N)\Delta t}e^{\boldsymbol{A}(t_{N-1})\Delta t}
	\dots e^{\boldsymbol{A}(t_1)\Delta t}e^{\boldsymbol{A}(t_0)\Delta t}\right)~.
\end{align}
This links the proposed technique to approximation methods based on
matrix-exponentials computed using discretized sampling \cite{Asslaender_CommunicationsPhysics_2019,Malik_Magn.Reson.Med._2020}.

This technique is not limited to the Bloch Equations,
but can be extended to also include the sensitivity 
analysis for the three partial derivatives
$R_1$, $R_2$ and $B_1$. This is further described in Appendix~\ref{sec::stm+sa}.

\subsection{Bloch Model-Based Reconstruction}

In the following, we integrate the generic Bloch operator $\mathcal{B}$ 
into a nonlinear model-based reconstruction framework with non-Cartesian,
calibrationsless, parallel imaging
and compressed sensing as illustrated in Figure~\ref{fig::overview}.
The reconstruction method solves the nonlinear inverse problem
for the maps $\boldsymbol{x}=(\boldsymbol{x_p}~\boldsymbol{x_c})^T$ with the physical
parameters $\boldsymbol{x_p} = \left(R_1~R_2~M_0~B_1\right)^T$ and coil sensitivities
$\boldsymbol{x_c}=(c_1~\dots~c_N)^T$ by optimizing
\begin{equation}
	\hat{\boldsymbol{x}} = \underset{\boldsymbol{x}}{\textrm{argmin}}\|\boldsymbol{y}-{\mathcal{A}}(\boldsymbol{x})\|_2^2+\alpha\boldsymbol{Q}(\boldsymbol{x_c})+\beta\boldsymbol{R}(\boldsymbol{x_p})~.
	\label{eq::optimization_NLINV}
\end{equation}
Eq. \ref{eq::optimization_NLINV} includes the forward-operator $\mathcal{A}$, the
measured data $\boldsymbol{y}$,
the Sobolev norm $\boldsymbol{Q}$ with its regularization parameter $\alpha$ to enforce
the smoothness of coil profiles \cite{Uecker_Magn.Reson.Med._2008} and $B_1$ maps.
A joint sparsity constraint $\boldsymbol{R}$ is applied to the other parameter
maps \cite{Vasanawala__2011,Beck_SIAMJ.Img.Sci._2009}.
The full forward operator is $\mathcal{A}=\mathcal{PFCB}$.
It is solved by the Iteratively Regularized Gauss-Newton Method (IRGNM)
\begin{align}
	\hat{\boldsymbol{x}}_{n+1} = \underset{\boldsymbol{x}}{\textrm{argmin}}\|D\mathcal{A}(\boldsymbol{x}_n)(\boldsymbol{x}-\boldsymbol{x}_n)+\mathcal{A}(\boldsymbol{x}_n)-\boldsymbol{y}\|_2^2\nonumber\\
	+\alpha_n\boldsymbol{Q}(\boldsymbol{x_c})+\beta_n\boldsymbol{R}(\boldsymbol{x_p})
	\label{eq::irgnm}
\end{align}
with the Jacobian $D\mathcal{A}(\boldsymbol{x}_n)$ and the 
regularization parameters $\alpha_n=\alpha_0\cdot q^n$ and
$\beta_n=\beta_0\cdot q^n$ at the $n$th iteration step.
Here, $\mathcal{C}$ is the nonlinear parallel imaging operator combining the signal
with the coil profiles, $\mathcal{F}$ represents the Fourier operator,
$\mathcal{P}$ the sampling pattern. The generic operator $\mathcal{B}$ takes information about the
applied sequence and outputs the simulated signal based on the STM technique. 
The partial derivatives of $\mathcal{B}$ are calculated using the direct sensitivity analysis.
The derivatives of $\mathcal{A}$ are described in Appendix~\ref{sec::frw_model_derivatives}.

\section{Methods}

\subsection{Implementation}

All simulations and reconstructions are implemented in the Berkeley Advanced
Reconstruction Toolbox (BART) using single-precision floating point arithmetic \cite{Uecker__2013}.
The Bloch operator $\mathcal{B}$ is implemented in BARTs nonlinear operator
framework \cite{Blumenthal_Magn.Reson.Med._2023}.
The calibrationless model-based reconstruction is based on an IRGNM-FISTA
following Wang \textit{et al.} \cite{Wang_Magn.Reson.Med._2018} and we refer to this work for further
details.
The Bloch operator includes a pixel-wise calculation of the signal evolution
using STMs (Section~\ref{sec:stm}) and of the partial derivatives
with SAB (Section \ref{sec::sa}).
ODEs are solved using the a Runge-Kutta algorithm (RK54) with adaptive step-sizes.
The error tolerances are chosen to be $10^{-7}$ for the simulation comparisons in Section
\ref{sec::sa} and \ref{sec:stm} as well as $10^{-6}$ for further reduced
computational costs in the Bloch model-based reconstructions.
The Runge-Kutta solver exploits weights published by Dormand and Prince \cite{Dormand_J.Comput.Appl.Math._1980}.
For balancing the relative scaling of the partial derivatives
during the optimization of Eq.~\ref{eq::optimization_NLINV} 
pre-conditioning following Wang \textit{et al.} \cite{Wang_Magn.Reson.Med._2018} is used.
The initial wavelet regularization is set to $\alpha_0$=$\beta_0$=1
and decreased by $q$=1/2 in each Newton iteration.
The output of the Bloch model operator $\mathcal{B}$ is scaled according
to Section~\ref{sec:simu_scaling}.
As a globalized Newton method, the IRGNM does not require fine
tuning of initial values. Here, the maps are initialized with the constants
$R_1=1$ Hz, $R_2=1$ Hz, $M_0=1$, $B_1=0$
and the coil profiles are initialized with zero.

For comparison, we also implemented the reparameterized Look-Locker model 
from equation \ref{eq::look-locker-rep} in the same model-based 
reconstruction framework following Wang \textit{et al.} \cite{Wang_Magn.Reson.Med._2018}.

\subsection{Validation of Bloch Simulation}

The accuracy of the SAB technique is validated
with an IR bSSFP sequence for tissue with $T_1/T_2=$ 1250$/$45 ms. 
An analytical solution for the IR bSSFP signal can be derived from the Bloch
equations \cite{Schmitt_Magn.Reson.Med._2004} assuming hard pulses, a perfect inversion
and a perfect non-selective excitation.
The symbolic partial derivatives with respect to the parameters $R_1$, $R_2$, $M_0$
and $B_1$ are calculated in Appendix~\ref{sec::IRbSSFP_symb_derivative} and
are used as ground truth.
For validation of the derivatives the ODE simulation parameters are chosen to 
be close to the assumptions of the analytical model.
The ODE solution for the derivatives is also compared to the difference quotient
techniques (DQ) calculated using two simulations $M_{t,p}$
and $M_{t,p+h}$ differing by a small perturbation $h$
for all parameters $p\in(T_1, T_2, M_0, B_1)$ and time points $t$:
\begin{equation}
	\frac{\partial\boldsymbol{M}_t}{\partial p} = \frac{M_{t,p+h} - M_{t,p}}{h}
\end{equation}
The size of the perturbation is decreased until numerical noise dominates.

To validate the STM approach in the presence of RF pulses, gradients, and relaxation,
a slice-selective excitation
of a Hamming-windowed sinc-shaped inversion pulse
with T$_{\text{RF}}=1$ ms, BWTP=1, $\Delta z=$10 mm, $G_z=10$ mT/m is simulated.
Relaxation parameters are selected based on typical human white matter values at 3 T: 
$T_1/T_2=$832$/$80 ms \cite{Wansapura_J.Magn.Reson.Imaging_1999}.
The STM simulation is computed with the Runga-Kutta solver and the magnetization just before the slice rewinder is
compared to a direct simulation exploiting the same solver.
For comparison a traditional Bloch simulation technique based on temporal discretization with rotational
matrices (ROT) is performed using a discretization rate of 1 MHz.
According to the analysis shown in Figure S3, 
a sampling rate of 1 MHz is required for ROT to accurately model complex spin dynamics 
that include slice-selective RF pulses.\\
The simulation speed is analyzed for a FLASH sequence with FA=8$^\circ$, TR$/$TE=3.1$/$1.7 ms and
T$_{\text{RF}}$=1 ms simulated for 101 isochromats homogeneously distributed along a slice of 0.02 m width
and using a slice-selection gradient of 12 mT/m. The tissue parameters are set to relaxation times 
of white matter at 3 T. The simulations were executed on a single 
Intel(R) i7-8565U CPU core at 1.80$~$GHz.

\subsection{Validation of Reconstruction}

To validate the model-based reconstruction we further perform validations on numerical
and experimental phantoms as well as in vivo data
for both single-shot IR FLASH and IR bSSFP sequences with
tiny golden-angle based radial sampling.
The IR bSSFP sequence includes a prior $\alpha/2$-TR/2 pulse to achieve a smooth signal
evolution during the transient state \cite{Deimling_ISMRM_1992,Deshpande_Magn.Reson.Med._2003}.
To be able to decouple the information of $T_1$ and $T_2$ for an IR bSSFP sequence
(see section \ref{sec:IR_bSSFP_decoupling}) a $B_1$ map is acquired
on the same slice using a vendor protocol based on 
rapid $B_1$ estimation with preconditioned RF pulses and 
Turbo-FLASH readout \cite{Chung_Magn.Reson.Med._2010}.
All sequence parameters are shown in Table~\ref{tab::seq-parameters}.

Phantom data for an IR FLASH sequence published by Wang et al. \cite{Wang_Magn.Reson.Med._2020} 
was downloaded from Zenodo \cite{wang2020}.
This data was measured on a 3 T Magnetom Skyra by Siemens
Healthcare (Erlangen, Germany) with a 20 channel head-coil.
The measured phantom is a commercial reference phantom
(Diagnostic Sonar LTD, Scotland, UK, Eurospin II, gel 3, 4, 7, 10, 14, and 16)
consisting of six tubes with known $T_1$ relaxation values surrounded by water.
A digital phantom
of the dataset is created with the same
sequence and acquisition characteristics as the downloaded measurement.
The relaxation parameters were set to the estimated reference $T_1$ and $T_2$
values from previous studies \cite{Wang_Magn.Reson.Med._2018,Sumpf_IEEETrans.Med.Imaging_2014}.
Additional phantom data for a radial single-shot IR bSSFP
sequence was acquired on the $T_2$ spheres of a NIST
phantom \cite{Stupic_Magn.Reson.Med._2021} on the same scanner and the same
20 channel head-coil.
For comparison gold-standard maps of $T_1$ and $T_2$ are estimated
on the same slice of the NIST phantom using fully-sampled single-echo
spin-echo sequences.
All phantom measurements were performed at constant room temperature ($\sim$21$^{\circ}$C).
The estimated quantitative values from gold-standard scans were used to
simulate a second digital phantom with the same geometry
as the measured T2-spheres of the NIST geometry (model 130).

Radial single-shot IR FLASH and IR bSSFP data for a brain of a healthy volunteer
was acquired with a 20 channel head-coil
on a Siemens Skyra 3T system (Siemens Healthcare, Erlangen, Germany)
after obtaining written informed consent.
The IR FLASH data was measured with a TR/TE of 4.1/2.58 ms,
a flip angle of 6$^{\circ}$, a band-width-time-product (BWTP) of 4
and a RF pulse duration of 1 ms.
The field-of-view (FOV) was 220x220 mm$^2$ measured
at 512 samples with two-fold oversampling.
The single-shot IR bSSFP data was acquired with two different pulse
durations: 1 ms and 2.5 ms respectively.
The repetition and echo times were set to 4.88/2.44 ms and
10.8/5.4 ms.
A FOV of 200x200 mm$^2$, a flip angle of 45$^{\circ}$,
a BWTP of 4 and a base resolution of 256
was chosen and the same for both measurements.

The phantom and in vivo IR FLASH datasets are reconstructed with both the
Look-Locker model and the Bloch model-based technique.
As the IR FLASH sequence is insensitive to $R_2$, its estimation 
in the Bloch model-based reconstruction was turned off by setting the scaling
in the pre-conditioner to zero.
The initial free decay of 15.3 ms during the non-selective hyperbolic secant
inversion pulse is corrected in both cases using the
correction published by Deichmann et al. \cite{Deichmann_Magn.Reson.Med._2005}.

The flexibility of the Bloch model-based reconstruction with its ability
to also model more complex sequences is demonstrated using
single-shot IR bSSFP.
In the analysis of the measured dataset the $B_1$ map is used in the
forward model for scaling the nominal flip angle
for each pixel. Additionally, the $B_1$ estimation is turned off by 
setting the scaling to zero in the pre-conditioning.\\
The method is numerically validated using a digital
phantom of the $T_2$ sphere of the NIST
phantom (model version 130) implemented in BART.
The multi-coil phantom is simulated in the frequency domain
with the same sequence parameters as the measurement.
The eight coils are compressed to four virtual coils using a
singular-value decomposition (SVD).
Complex Gaussian noise is added before coil compression to further avoid an inverse crime.
To ensure realistic physical conditions the simulated signal model includes
a non-selective hyperbolic
secant inversion pulse and slice-selective excitations using multiple isochromats
distributed over equally spaced slice-selection gradient positions.\\
While analytical solutions of the Bloch equations require the assumption of
perfect inversions and ideal non-selective excitation, the generic simulation of the 
Bloch model-based reconstruction technique can simulate more realistic signal models.
We show how step-wise improvements to the model allow more accurate
modelling of actual measurements. This is demonstrated on the
numerical and measured NIST phantom
datasets by performing reconstructions with various different assumptions 
about the signal model.
The first reconstruction uses a model close to the analytical formula
by assuming a perfect inversion and a non-selective excitation.
We then add a realistic slice-selective excitation simulated as mean signal of various
homogeneously spaced isochromats along the slice-selection gradient.
To also model the effect of non-optimal inversion
efficiency, the final reconstruction includes an extended model with
realistic non-selective hyperbolic secant inversion.

The radial single-shot IR bSSFP in vivo data is reconstructed using
the most realistic model.
For comparison, the single-shot IR FLASH measurement was acquired on the same slice 
and reconstructed with the Bloch model-based reconstruction assuming a realistic
IR FLASH signal model with non-selective
hyperbolic secant inversion and gradient
based slice-selective excitation model to estimate a $T_1$ map.

Details about the measurements can be found in Table \ref{tab::seq-parameters}.

\section{Results}

\subsection{Validation of Bloch Simulation}

Figure \ref{fig::sab}
shows the partial derivatives for an IR bSSFP sequence with respect to $R_1$,
$R_2$ and $B_1$ for the analytical reference, the SAB technique and difference quotient
(DQ) techniques with different perturbations $h$ on the left.
On the right Figure \ref{fig::sab} presents the differences of DQ and SAB to the
analytical reference.\\
As expected, the error of DQ decreases for small perturbations until
numerical noise starts to dominate for very small $h$.\\

The SAB technique demonstrates a high accuracy and precision of
estimating partial derivatives without requiring tuning of the perturbation level.

Figure \ref{fig::stm}.A compares the simulation results of a Hamming-windowed
sinc-shaped inversion pulse using the Runge-Kutta 54 method with Dormand-Prince weights (RK54) \cite{Dormand_J.Comput.Appl.Math._1980},
STM, and ROT technique.

The error of the STM simulation is dominated by numerical noise 
due to limited floating point precision. 
With the parameters used here, the STM technique
has substantially lower point-wise errors than ROT.\\

It demonstrates that STM reproduces the RK54 technique for finding solutions to the
Bloch equations extremely well, while ROT is affected by errors due to the
discretization with fixed sampling rate and its nature of being a
first order method constrained to single floating point precision here.
The runtime of RK54, STM and ROT is shown in Figure \ref{fig::stm}.B.
The computational cost of ROT increases linear with higher sampling rates.
The STM has higher initial costs than the other techniques which 
reflect the initial calculation of the state-transition matrices.
The other methods are therefore faster for a small number of repetitions.
For more repetitions, the STM becomes much faster as it 
requires only a few matrix multiplications per TR.
A detailed comparison of the computational cost and accuracy of the RK54, STM and ROT techniques
for various error tolerances and sampling rates can be found in Supplementary Section S3 
and Supplementary Figure S3. 

\subsection{Validation of Reconstruction}

The Bloch model-based reconstruction was compared to
the Look-Locker model-based version for simulated (Figure \ref{fig:irflash}.A) and measured
single-shot IR FLASH phantom data (Figure \ref{fig:irflash}.B).
Both methods recover high quality $T_1$ maps with
small differences.
Values for the same Regions-of-Interests (ROIs) are 
very similar leading to their position on the diagonal of the Bloch vs. 
Look-Locker plot on the most right of Figure \ref{fig:irflash}.A
and Figure \ref{fig:irflash}.B.
The reconstructed tubes are very homogeneous in both reconstructions
leading to low standard deviations.
Reconstructions using different regularization parameters or
no regularization are shown in Supplementary Figure S4 
and Supplementary Figure S6, 
respectively.

In the difference map between the $T_1$ maps of the two methods only the water background shows areas with minor differences.
This probably results from small differences in the Sobolev regularization on the flip angle map in both techniques.
At the walls of the inner tubes there is not enough signal and the $T_1$ maps are not well defined.
Results for the radial single-shot in vivo IR FLASH data are
shown in Figure \ref{fig:irflash}.C.
The parameter maps are visually indistinguishable
except for minor artifacts in the areas of the head with flow related effects, which are not modelled by both signal models.
The $T_1$ values in the marked ROIs for representative white and gray matter areas show a
very good correspondence. The homogeneity within the white matter
is high corresponding to a small standard deviation.

The complex $M_0$ parameter map in Figure \ref{fig:irflash}.D reconstructed with the Bloch model-based reconstruction is of good quality showing no artifacts and a
homogeneous phase. Only in border regions phase changes
are present which are most likely caused by fat.
The relative flip angle map is globally lower than one.
It combines the effect of an imperfect slice-selective excitation and the present $B_1$ field.
The intensity and phase of the estimated SVD-compressed virtual coil sensitivities
is comparable to intensity and phase of sensitivities estimated
with ESPIRiT \cite{Uecker_Magn.Reson.Med._2014}(results not shown).

Figure \ref{fig:irbssfp_nist} shows the reconstructed $T_1$ and $T_2$ maps 
of the digital NIST phantom (\ref{fig:irbssfp_nist}.A) and the
measurement (\ref{fig:irbssfp_nist}.B) using the radial single-shot IR bSSFP acquisition
for different models compared to the reference values in Bland-Altman plots.
The most simplistic model assumes a perfect inversion and an
ideal non-selective excitation (\textbf{Perfect Inversion}) and shows 
inaccuracies in the $T_1$ and $T_2$ estimation.
By integrating a slice-selective excitation (\textbf{Slice}) the errors in $T_2$ are significantly reduced
leaving an offset in $T_1$.
Adding a realistic hyperbolic secant inversion pulse to the forward model
(\textbf{Pulse+Slice}) corrects for the $T_1$ offset
leading to an accurate estimation of the relaxation parameters.
These effects are present in both: the simulation and the measured data reconstructions.
Important to note is that the NIST phantoms
contains some spheres with extreme  $T_1$ and $T_2$ parameters \cite{Stupic_Magn.Reson.Med._2021},
which were excluded from the Bland-Altman analysis
in Figure \ref{fig:irbssfp_nist}.A and \ref{fig:irbssfp_nist}.B for improved visualization.
In particular, the three highest $T_2$ values (1.450 s, 0.388 s and 0.271 s)
were removed for the reconstruction from simulated data and the highest
and lowest $T_2$ values (1.450 s, 0.006 s) for the measured data.
Especially the simple model has difficulties
in finding the correct relaxation parameters in
the reconstruction of the measured data, so that the mean
value is outside the plotted region.
A direct comparison of reference and estimated parameters
in a diagonal plot can be found in Supplementary Figure S1. 
A Bland-Altman plot with all data points is shown in Supplementary Figure S2. 

Reconstructed $T_1$ and $T_2$ parameter maps for the
two single-shot IR bSSFP scans of a human brain
are shown in Figure \ref{fig:irbssfp_invivo}.
For comparison, a map with the Look-Locker model-based reconstruction 
of the IR FLASH scan of the same slice is added. Both IR bSSFP
reconstructions show large offsets in $T_1$ compared to the IR FLASH reference.
The relaxation values for the two analyzed ROIs are listed in Table \ref{tab::invivo-irbssfp}.
The differences are smaller for the longer TR and longer RF pulse duration $T_{\text{RF}}$ compared
to the short pulse protocol.
This is likely due to the magnetization transfer effect (MT)
that affects the IR bSSFP sequence but is not included in the current study.

The reconstruction times depend on the complexity of the forward model.
The reconstruction of the IR FLASH datasets took about 80 s on a AMD EPYC 7662 64-Core CPU
and a Nvidia A100-SXM-80GB GPU. The reconstruction of the
simple forward model of the IR bSSFP NIST
phantom dataset took 60 s, while the most complex model reconstruction took 38 min.
The longest reconstruction times were required for the in vivo IR bSSFP dataset with short RF pulse.
The strong slice-selection gradient prolonged the reconstruction to about 75 min.

\section{Discussion}

A nonlinear model-based reconstruction framework can be used
in combination with well-crafted sequences and their analytical
signal representations to accelerate quantitative MRI.
This work presents a generalization of this well-known
approach to arbitrary MRI sequences by exploiting
the Bloch equations directly as forward
model. This method becomes computationally feasible by including
a direct sensitivity analysis of the Bloch equations. It allows us to
use a generic ODE solver to compute the derivatives required
by efficient nonlinear optimization algorithms such as the IRGNM.
In comparison to techniques based on difference quotients it
produces highly stable and accurate partial derivatives without
the need of fine tuning perturbation levels.
This was shown by estimating partial
derivatives of an IR bSSFP experiment with the mentioned techniques
and by comparing their results to the underlying
analytical reference.

To further reduce computational demand, we exploit 
pre-computed STMs.
They are used to solve the Bloch equations and the system
describing their sensitivities simultaneously
for arbitrary initial conditions for a given time span.
They reduce the spin dynamics even in the presence
of external fields, gradients, and relaxation
to single matrix multiplications.
It dramatically speeds up the reconstruction
whenever the MRI sequence contains
repeated patterns as it is often the case.
In the presented example of the FLASH sequence with
101 isochromats along a slice and
simulated for 1000 repetitions,
the run-time of the simulation
was reduced by a factor of 10 from 5 s down to 0.5 s
in comparison to a regular Runge-Kutta ODE solver.
Even in the presence of gradients, RF pulses and relaxation
the slice-profile analysis showed the high accuracy of the
STM technique in reproducing the ODE solver results.

Experimentally we confirmed that the Bloch model-based
reconstruction reproduces the Look-Locker model as a special case.
A comparison between both techniques showed only minor differences in the $T_1$ maps
reconstructed from the single-shot phantoms and single-shot in vivo data.\\
The integration of a generic Bloch simulation into the reconstruction adds
the flexibility to analyze a broad variety of sequences.
As an initial example, we applied the technique to IR bSSFP sequence
and validated it using a numerical and measured NIST phantom dataset.
By correctly modelling
the slice-selective excitation and a non-selective
hyperbolic secant inversion pulse highly accurate
$T_1$ and $T_2$ maps could be obtained.

For a human brain, $T_1$ maps estimated from an IR bSSFP sequence
were compared to Bloch model-based reconstructions
of an IR FLASH acquisition of the same slice both including non-selective
hyperbolic inversion and a slice-selective excitation.
Here, differences could be observed which are likely caused by MT \cite{Bieri_Magn.Reson.Med._2006}.
This hypothesis is supported by the fact that prolonging the RF pulse duration and 
increasing the TR reduced the differences, but preliminary
results (Supplementary Section S5) 
suggest that this does
not explain the complete discrepancy and that other effects may
also play a role. The NIST phantom measurement is not affected by MT effects,
because it is based on water \cite{Stupic_Magn.Reson.Med._2021}.
The IR FLASH measurement is assumed to be unaffected by MT because of its small flip angle 
\cite{Graham_J.Magn.Reson.Imaging_1997}.

At this stage, the most relevant practical limitation is the need
to manually tune the scaling factors used for pre-conditioning.
For each analyzed sequence the relative scaling between the partial derivatives needs
to be balanced manually to ensure smooth convergence.
Future work is going to investigate automatic scaling techniques
\cite{Maier_Magn.Reson.Med._2018,Tan_NMRBiomed._2017}.

Further extensions could be the application to hybrid state free precession sequences \cite{Asslaender_CommunicationsPhysics_2019,Asslaender_Magn.Reson.Med._2019},
multi-echo inversion-recovery sequences \cite{Feng_Magn.Reson.Med._2021}, and
magnetization transfer models \cite{Malik_Magn.Reson.Med._2020}.

\section{Conclusion}

This work developed a generic framework for model-based reconstruction
using the Bloch equations. The approach is validated 
numerically and tested experimentally using phantom and
in-vivo scans.

\section*{Acknowledgments}
We thank Moritz Blumenthal and Christian Holme
for discussions and help with BART.

This project was supported by the DZHK (German Centre for Cardiovascular Research),
and funded in part by the Deutsche Forschungsgemeinschaft (DFG, German Research Foundation)
under Germany’s Excellence Strategy - EXC 2067/1- 390729940,
and funded in part by NIH under grant U24EB029240.
We also gratefully acknowledge the support of the NVIDIA corporation with the
donation of one NVIDIA TITAN Xp GPU for this research.

\subsection*{Data Availability}

The data of this work was uploaded to \href{https://zenodo.org/record/6992763}{Zenodo @doi:10.5281/zenodo.6992763}.\\
The scripts reproducing all figures of this manuscript are published at \href{https://github.com/mrirecon/bloch-moba}{Github @mrirecon/bloch-moba}.\\
The reconstruction code is implemented in  \href{https://github.com/mrirecon/bart}{BART} with commit \texttt{e641e74b}.\\
A tutorial about the usage of the Bloch model-based reconstruction with BART can be found at \href{https://github.com/mrirecon/bloch-tutorial}{Github @mrirecon/bloch-tutorial}.

\subsection*{Conflict of interest}

The authors declare no potential conflict of interests.

\bibliographystyle{MRM-AMA}
\bibliography{../radiology.bib}%
\vfill\pagebreak

\paragraph{List of Figures}

\textbf{Figure \ref{fig::overview}} Illustration of the operators used in the Bloch model-based reconstruction.
The bottom part presents the ODEs for the signal and the derivatives.\\
\textbf{Figure \ref{fig::sab}} \textbf{Left}: Temporal evolution of the partial derivatives
with respect to $R_1$, $R_2$ and $B_1$ estimated for an
IR bSSFP sequence with the SAB, DQ with varying perturbation
levels $h$ and the analytical references.
\textbf{Right}: Plot with point-wise errors of the various DQ methods and the
SAB with respect to the analytical reference.
Note that the errors are presented in ppm for visualization.\\
\textbf{Figure \ref{fig::stm}} \textbf{A}: The slice-selection gradient based simulation
for a Hamming-windowed sinc-shaped
inversion pulse simulated with the RK54 framework is shown (left).
The point-wise errors of the RK54 (top), the STM technique (center) and the ROT method (bottom) with
sampling rate 1 MHz are plotted for the x-, y- and
z- component of the magnetization.
Note that the errors are scaled by large factors for visualization.
\textbf{B}: The runtime of the STM technique is compared to the
reference RK54 method and a ROT simulation performed with a sampling rate of 1 MHz.
The simulation is performed for 101 isochromats
homogeneously distributed along a slice-selection gradient during a
FLASH sequence for various numbers of repetitions.
The maximum of 1000 is chosen to cover about 4 s of acquisition, required to
measure enough data points for mapping high $T_1$ values.
A more detailed version of this figure has been added to the Supplementary Section S3.\\
\textbf{Figure \ref{fig:irflash}} Reconstructed $T_1$ parameter maps for radial single-shot IR FLASH data
acquired from a numerical (\textbf{A}) and measured phantom (\textbf{B}) as well as a human brain (\textbf{C}).
The Bloch model-based reconstruction and the differences between the
two methods are shown in the middle. The difference map is scaled up 
by a factor of 20 to improve visualization.
On the right the $T_1$ values of the color-coded ROIs (arrows) 
of the Bloch reconstruction vs. Look-Locker reconstruction
are plotted together with standard deviations.
Besides the in vivo $T_1$ map presented in \textbf{C}, the Bloch model-based technique
reconstructs a complex valued $M_0$ map, a relative flip angle map and
complex coil sensitivities shown in \textbf{D}.\\
\textbf{Figure \ref{fig:irbssfp_nist}} \textbf{A}: Reconstructed $T_1$ and $T_2$ parameter maps and the corresponding
ROI values for numerical radial single-shot IR bSSFP data
of a digital multi-coil reference object simulated in k-space.
The left side shows the reconstructed parameter maps and the right the ROI analysis
results in Bland-Altman plots relative to the simulated reference values.
The analyzed ROIs are marked and numbered in the $T_1$ map.\\
\textbf{B}: Reconstructed $T_1$ and $T_2$ parameter maps and the corresponding
ROI values for a radial single-shot IR bSSFP measurement of the $T_2$ spheres
of the NIST system phantom.
The left side corresponds to the rightmost ROI analysis.
The right side presents the comparison of the analyzed ROIs
reconstructed with the Bloch model-based technique
for various signal model assumptions
compared with gold-standard reference values in Bland-Altman plots.
For improved visualization individual outliers are removed from the plot.\\
\textbf{Figure \ref{fig:irbssfp_invivo}} The $T_1$ parameter map reconstructed from a radial
single-shot IR FLASH in vivo dataset with a Look-Locker model-based reconstruction
is shown on the left.
It also shows the reconstructed $T_1$ parameter maps of a radial
single-shot IR bSSFP in vivo dataset acquired on the same brain slice
for short RF pulses ($D_{\text{RF}}$: 1 ms, TR: 4.88 ms) on the top and
long RF pulses ($D_{\text{RF}}$: 2.5 ms, TR: 10.8 ms) on the bottom
reconstructed with the Bloch model-based reconstruction.
In the center column the difference maps are shown.
The values corresponding to the colored ROIs are listed in Table \ref{tab::invivo-irbssfp}.
On the right the $T_2$ parameter maps for the short and long RF pulse experiments
are shown reconstructed from the IR bSSFP sequence.\\
\textbf{Figure S1} \textbf{A}: Reconstructed $T_1$ and $T_2$ parameter maps and the corresponding
ROI values for numerical radial single-shot IR bSSFP data
of a digital multi-coil reference object simulated in k-space.
The left side shows the reconstructed parameter maps and the right the ROI analysis
results relative to the simulated reference values wit standard derivations.
The analyzed ROIs are marked and numbered in the $T_1$ map.
\textbf{B}: Reconstructed $T_1$ and $T_2$ parameter maps and the corresponding
ROI values for a radial single-shot IR bSSFP measurement of the $T_2$ spheres
of the NIST system phantom.
The left side corresponds to the rightmost ROI analysis.
The right side presents the comparison of the analyzed ROIs
reconstructed with the Bloch model-based technique
for various signal model assumptions
compared with gold-standard reference values.\\
\textbf{Figure S2} \textbf{A}: Reconstructed $T_1$ and $T_2$ parameter maps and all corresponding
ROI values for simulated radial single-shot IR bSSFP data.
Here, the datapoints removed to improve visualization in Figure \ref{fig:irbssfp_nist}
are included. Reconstructed parameter maps are shown on the left
Bland-Altman plots using the simulated reference values are shown on the right.
The ROIs are marked and numbered in the $T_1$ map.
\textbf{B}: Reconstructed $T_1$ and $T_2$ parameter maps and the corresponding
ROI values for a radial single-shot IR bSSFP measurement of the $T_2$ spheres
of the NIST system phantom.
The left side corresponds to the rightmost ROI analysis.
The right side shows the Bland-Altman plots comparing the
values of the ROIs reconstructed with the Bloch model-based technique
for various assumptions used in the signal model
compared with gold-standard reference values.
For improved visualization individual outliers were removed from the plot.\\
\textbf{Figure S3} \textbf{A}: $T_1$ maps reconstructed from
single-shot IR FLASH (same as for Figure \ref{fig:irflash}.B) for varying
tolerance for computation of STMs. The forward model assumes on-resonant spins.
The difference of the reconstructions for tolerances of 1e-5 to 1e-2 compared
to the 1e-6 reference are shown in the lower row. The differences are scaled
by larges factors for improved visualization.\\
\textbf{B}: Same analysis as for part A, but assuming a complex forward model
including a slice-selection gradient.\\
\textbf{C}: Extension of Figure \ref{fig::stm}.A with varying tolerances
for RK54 and STM and sampling rates for ROT.\\
\textbf{D}: Extension of Figure \ref{fig::stm}.B with varying tolerances
for RK54 and STM as sampling rates for ROT.\\
\textbf{Figure S4} Reconstructed $T_1$ parameter maps similar to Figure \ref{fig:irflash}.B for varying
wavelet regularization strengths $\beta_0$ for the Bloch model-based reconstruction (upper row).
On the left a reference reconstruction estimated with
the re-parameterized Look-Locker model-based technique is shown.
In the bottom row the differences between Bloch model-based reconstruction
and the reference map are shown scaled by a factor of 10 for improved
visualization.\\
\textbf{Figure S5} Visualization of the magnetization transfer effect in $T_1$ maps
reconstructed from an IR bSSFP sequence with varying TR and $T_{RF}$.
On the left the $T_1$ map corresponding to the longest $T_{RF}=2.5$ ms is plotted with colored ROIs. The mean value of the relaxation parameter values for the colored areas is plotted with its standard deviation on the right for various length of RF-pulses $T_{RF}$. The dotted lines represent the fitted values following the analysis of section S5.\\ 
\textbf{Figure S6} Bloch model-based reconstruction of the short RF pulse
in vivo single-shot IR bSSFP dataset. The coil-sensitivities have been
estimated with the regular method and have been added as fixed prior knowledge
to both reconstruction. The top presents the regularized reconstruction,
while the bottom results do not include any regularization on the
parameter maps.

\paragraph{List of Tables}

\textbf{Table \ref{tab::seq-parameters}} Table listing the sequence parameters for the performed measurements of this work.\\
\textbf{Table \ref{tab::invivo-irbssfp}} Table listing the single-shot in-vivo IR bSSFP ROI analysis results presented in Figure \ref{fig:irbssfp_invivo}.\\
\textbf{Table S1} Table listing the fitting parameters estimated for Figure S5.\\ 
\textbf{Table S2} Table listing the sequence parameters for the analysis in Figure S5. 

\appendix

\section{Combining Sensitivity Analysis with a Direct Sensitivity Analysis}
\label{sec::stm+sa}

The system matrix $\boldsymbol{A}(t)$ in
equation \ref{eq::Bloch_system_matrix} can be extended
to include the sensitivity analysis for the three partial
derivatives $R_1$, $R_2$ and $B_1$:
\setcounter{MaxMatrixCols}{20}

\noindent
\makebox[\textwidth]{\parbox{1.7\textwidth}{%
\begin{equation}
	\boldsymbol{A}(t) =
	\scalebox{.39}{$
		\begin{pmatrix}
			-R_2							& 	\gamma B_z(t)						&	-\gamma \sin(\phi(t))B_1(t) B_z(t)
			& 0& 0&	0& 0& 0& 0& 0& 0& 0& 0\\
			-\gamma B_z(t)						&	-R_2							&	\gamma \cos(\phi(t))B_1(t) B_z
			& 0& 0&	0& 0& 0& 0& 0& 0& 0& 0\\
			\gamma \sin(\phi(t))B_1(t) B_z(t)	&	-\gamma \cos(\phi(t))B_1(t) B_z(t)	&	-R_1
			& 0& 0&	0& 0& 0& 0& 0& 0& 0& M_0 R_1\\
			0& 0& 0
			&	-R_2						&	\gamma B_z(t)						&	-\gamma \sin(\phi(t))B_1(t) B_z(t)
			& 0& 0& 0& 0& 0& 0& 0\\
			0& 0& 0
			&	-\gamma B_z(t)					&	-R_2							&	\gamma \cos(\phi(t))B_1(t) B_z(t)
			& 0& 0& 0& 0& 0& 0& 0\\
			0& 0& -1
			&	\gamma \sin(\phi(t))B_1(t) B_z(t)&	-\gamma \cos(\phi(t))B_1(t) B_z(t)	&	-R_1
			& 0& 0& 0& 0& 0& 0&	M_0\\
			-1& 0& 0& 0& 0&	0
			& -R_2 							&	\gamma B_z(t)						&	-\gamma \sin(\phi(t))B_1(t) B_z(t)
			& 0& 0&	0&	0\\
			0 & -1 & 0& 0& 0& 0
			& -\gamma B_z(t)					&	-R_2							&	\gamma \cos(\phi(t))B_1(t) B_z(t)
			& 0& 0&	0&	0\\
			0& 0& 0& 0& 0& 0
			&\gamma\sin(\phi(t))B_1(t) B_z(t)	&	-\gamma \cos(\phi(t))B_1(t) B_z(t)	&	-R_1
			& 0& 0& 0& 0\\
			0& 0& -\gamma\sin(\phi(t))B_1(t) & 0& 0& 0& 0& 0& 0
			&	-R_2						&	\gamma B_z(t)						&	-\gamma \sin(\phi(t))B_1(t) B_z(t)
			& 0\\
			0& 0& \gamma\cos(\phi(t))B_1(t) & 0& 0& 0& 0& 0& 0
			& -\gamma B_z(t)					&	-R_2							&	\gamma \cos(\phi(t))B_1(t) B_z(t)
			& 0\\
			\gamma\sin(\phi(t))B_1(t)& -\gamma\cos(\phi(t))B_1(t)& 0& 0& 0& 0& 0& 0& 0
			& \gamma \sin(\phi(t))B_1(t) B_z(t)	&	-\gamma \cos(\phi(t))B_1(t) B_z(t)	&	-R_1
			& 0\\
			0& 0& 0& 0& 0& 0& 0& 0& 0& 0& 0& 0&	0
		\end{pmatrix}
		$}.
\end{equation}
}}

with its corresponding parameter vector from equation \ref{eq::bloch_homo}:
\begin{align}
	\boldsymbol{M}(t)\rightarrow\boldsymbol{x}(t)=
	\begin{pmatrix}
		M_x(t)\\
		M_y(t)\\
		M_z(t)\\
		Z_{R_1,x}(t)\\
		Z_{R_1,y}(t)\\
		Z_{R_1,z}(t)\\
		Z_{R_2,x}(t)\\
		Z_{R_2,y}(t)\\
		Z_{R_2,z}(t)\\
		Z_{B_1,x}(t)\\
		Z_{B_1,y}(t)\\
		Z_{B_1,z}(t)\\
		1
	\end{pmatrix}.
\end{align}

\section{Forward Model Derivatives}
\label{sec::frw_model_derivatives}
The derivative of $\mathcal{A}$ in equation \ref{eq::optimization_NLINV}
follows by exploiting the Jacobi matrix and the product rule
similar to \cite{Uecker_Magn.Reson.Med._2008,Wang_Magn.Reson.Med._2018}.

\noindent
\makebox[\textwidth]{\parbox{1.3\textwidth}{%
\begin{equation}
	\boldsymbol{D}\mathcal{A}(\boldsymbol{x})
	\begin{pmatrix}
		\textrm{d}R_1\\
		\textrm{d}R_2\\
		\textrm{d}M_0\\
		\textrm{d}B_1\\
		\textrm{d}c_1\\
		\vdots\\
		\textrm{d}c_N
	\end{pmatrix}
	= 
	\begin{pmatrix}
		\mathcal{P}\mathcal{F}\left(\textrm{d}c_1 M_{t_1} + c_1\left[
		\frac{\partial M_{t_1}}{\partial R_1}\textrm{d}R_1+
		\frac{\partial M_{t_1}}{\partial R_2}\textrm{d}R_2+
		\frac{\partial M_{t_1}}{\partial M_0}\textrm{d}M_0+
		\frac{\partial M_{t_1}}{\partial B_1}\textrm{d}B_1
		\right]\right)\\
		\vdots\\
		\mathcal{P}\mathcal{F}\left(\textrm{d}c_N M_{t_1} + c_N\left[
		\frac{\partial M_{t_1}}{\partial R_1}\textrm{d}R_1+
		\frac{\partial M_{t_1}}{\partial R_2}\textrm{d}R_2+
		\frac{\partial M_{t_1}}{\partial M_0}\textrm{d}M_0+
		\frac{\partial M_{t_1}}{\partial B_1}\textrm{d}B_1
		\right]\right)\\
		\vdots\\
		\mathcal{P}\mathcal{F}\left(\textrm{d}c_N M_{t_n} + c_N\left[
		\frac{\partial M_{t_n}}{\partial R_1}\textrm{d}R_1+
		\frac{\partial M_{t_n}}{\partial R_2}\textrm{d}R_2+
		\frac{\partial M_{t_n}}{\partial M_0}\textrm{d}M_0+
		\frac{\partial M_{t_n}}{\partial B_1}\textrm{d}B_1
		\right]\right)
	\end{pmatrix}.
	\label{eq::derivative of forward_op}
\end{equation}
}}

The adjoint of the derivative becomes
\begin{equation}
	\boldsymbol{D}\mathcal{A}^H(\boldsymbol{x})
	\begin{pmatrix}
		y_{1,1}\\
		y_{2,1}\\
		\vdots\\
		y_{n,N}
	\end{pmatrix}
	=
	\begin{pmatrix}
		\textrm{d}R_1\\
		\textrm{d}R_2\\
		\textrm{d}M_0\\
		\textrm{d}B_1\\
		\textrm{d}c_1\\
		\vdots\\
		\textrm{d}c_N
	\end{pmatrix}
	=
	\begin{pmatrix}
		\sum\limits_{j=1}^N\sum\limits_{k=1}^n
		\overline{\left(\frac{\partial M_{t_k}}{\partial R_1}\right)}
		\cdot\overline{c_j}\cdot
		\mathcal{F}^{-1}\left[\mathcal{P}^H y_{k,j}\right]\\
		\sum\limits_{j=1}^N\sum\limits_{k=1}^n
		\overline{\left(\frac{\partial M_{t_k}}{\partial R_2}\right)}
		\cdot\overline{c_j}\cdot
		\mathcal{F}^{-1}\left[\mathcal{P}^H y_{k,j}\right]\\
		\sum\limits_{j=1}^N\sum\limits_{k=1}^n
		\overline{\left(\frac{\partial M_{t_k}}{\partial M_0}\right)}
		\cdot\overline{c_j}\cdot
		\mathcal{F}^{-1}\left[\mathcal{P}^H y_{k,j}\right]\\
		\sum\limits_{j=1}^N\sum\limits_{k=1}^n
		\overline{\left(\frac{\partial M_{t_k}}{\partial B_1}\right)}
		\cdot\overline{c_j}\cdot
		\mathcal{F}^{-1}\left[\mathcal{P}^H y_{k,j}\right]\\
		\sum\limits_{k=1}^n\overline{M_{t_k}}\cdot\mathcal{F}^{-1}\left[\mathcal{P}^H y_{k,1}\right]\\
		\vdots\\
		\sum\limits_{k=1}^n\overline{M_{t_k}}\cdot\mathcal{F}^{-1}\left[\mathcal{P}^H y_{k,N}\right]
	\end{pmatrix}.
	\label{eq::adjoint_derivative of forward_op}
\end{equation}

\section{Look-Locker Reparameterization}
\label{sec::LL_repara}

The Look-Locker model represents a special solution of the Bloch equations
for an IR FLASH sequence \cite{Look_Rev.Sci.Instrum._1970}.
It assumes a perfect inversion, a small flip angle and short repetition times
compared to the relaxation effects.
Initial relaxation effects between the inversion and the first echo
can be compensated analytically \cite{Deichmann_Magn.Reson.Med._2005}.
The original formulation of the Look-Locker model with the
parameters $M_0$, $M_{ss}$ and $R_1^*$ is
\begin{equation}
	M_z(M_{ss},M_0,R_1^*,t) = M_{ss}-(M_{ss}+M_0)\cdot e^{-R_1^*\cdot t}~.
	\label{eq::look-locker}
\end{equation}
These parameters are related to the underlying physical
parameters $M_0$, $R_1$ and $\alpha_{\text{eff}}$.
With the assumption of short repetition times \cite{Deichmann_J.Magn.Reson._1992} 
\begin{equation}
	\frac{M_0}{M_{ss}}=\frac{R_1^*}{R_1}
\end{equation}
the effective relaxation rate
\begin{equation}
	R_1^*=R_1+R'_1=R_1-\frac{1}{\text{TR}}\ln\cos\alpha_{\text{eff}}
\end{equation}
splits into $R_1$ and the readout relaxation rate $R'_1$ determined by
the effective flip angle $\alpha_{\text{eff}}$ \cite{Roeloffs_Int.J.Imag.Syst.Tech._2016}.
The reparameterized Look-Locker model can be formulated as
\begin{equation}
	\footnotesize
	M_z(M_0, R_1, R'_1,t)=M_0\cdot\left(\frac{R_1}{R_1+R'_1}-\left(1+\frac{R_1}{R_1+R'_1}\right)\cdot e^{-(R_1+R_1')t}\right).
	\label{eq::look-locker-rep}
\end{equation}
Here, Equation~\ref{eq::look-locker-rep} depends directly on the physical parameters
$M_0$, $R_1$ and $R'_1=-\frac{1}{\text{TR}}\ln\cos\alpha_{\text{eff}}$
and allows adding prior knowledge
about smooth B1 maps \cite{Clare_Magn.Reson.Med._2001,Deichmann_Magn.Reson.Med._2005} to the
reconstruction \cite{Roeloffs_Int.J.Imag.Syst.Tech._2016}.
This reparameterized model can be directly compared to a Bloch model-based reconstruction
using the same set of parameters.

\section{Scaling Factors}
\label{sec:simu_scaling}

The output of the forward operator $\mathcal{B}$ in equation \ref{eq::optimization_NLINV}
is the strength of the signal estimated by the Bloch simulation and its partial
derivatives.
The strength of the signal depends on the sequence and especially the applied flip angle.
Therefore, the signals output differs for classical FLASH or bSSFP sequences
influencing the weighting between the data fidelity and regularization terms
changing the optimization behaviour.

A generic reconstruction aims for robustness against variations of the sequence parameters.
It requires a scaling of the signal and its partial derivatives simulated within $\mathcal{B}$.
The implemented scaling is motivated by the Look-Locker model assumption
that the longitudinal magnetization $M_z$
in equation \ref{eq::look-locker-rep} is proportional to the measured signal $M_{xy}$
and scaled to 1.
Thus, the signal of a simulation with an initial
magnetization of length 1 requires scaling of the
simulated signal output $M_{xy}$ by the applied flip angle $\alpha$ and the relaxation
effect $e^{-\frac{t_{\text{TE}}}{T_2}}$ during the echo time interval $\Delta t_{\text{TE}}$:

\begin{equation}
	M_z = \frac{e^{-\frac{\Delta t_{\text{TE}}}{T_2}}}{\sin\alpha}\cdot M_{xy}~.
	\label{eq::IRFLASH_scaling}
\end{equation}

Because of the short echo times of the sequences used
in this work, the $T_2$ relaxation effect can be neglected.
This assumption avoids additional $T_2$ dependencies of the estimated
derivatives.
The final scaling factor $\frac{1}{\sin\alpha}$
increases the robustness of the forward operator $\mathcal{B}$
in equation \ref{eq::optimization_NLINV} to the choice of the applied flip angle
in FLASH based sequences.
For a bSSFP type sequence the flip angle in equation \ref{eq::IRFLASH_scaling}
needs to be halved to take its dynamics on the $\alpha/2$ cone into
account \cite{Scheffler_Eur.Radiol._2003}.

\section{IR bSSFP Information Encoding}
\label{sec:IR_bSSFP_decoupling}

The IR bSSFP signal behaviour is described by a limited exponential growth similar
to the IR FLASH sequence \cite{Schmitt_Magn.Reson.Med._2004}.
A single inversion recovery can encode information for estimating 3 parameters.
While the IR FLASH sequence is sensitive to exactly 3 parameters, the IR bSSFP
is also sensitive to $T_2$, leading to 4 parameters in total.
With a single limited exponential growth this additional parameter can not be encoded
and two parameters need to be coupled.
For bSSFP sequences the relaxation parameters $T_1$ and $T_2$ are coupled.
Prior knowledge about $B_1$ can be used to decouple both relaxation parameters \cite{Schmitt_Magn.Reson.Med._2004,Ehses_Magn.Reson.Med._2013,Pflister_Magn.Reson.Med._2019}.

\section{Symbolic Derivatives of IR bSSFP}
\label{sec::IRbSSFP_symb_derivative}

The analytical signal model for an IR bSSFP can be derived from the Bloch equations
with the assumptions of hard RF pulses, a perfect inversion and an ideal $\alpha/2-\text{TR}/2$
magnetization preparation.

The signal is modelled by \cite{Schmitt_Magn.Reson.Med._2004}:
\begin{equation}
	M(M_{ss}, M_0^*,R_1^*,t)=M_{ss}-(M_0^*+M_{ss})\cdot e^{-R_1^*\cdot t}
	\label{eq:irbssfp_signal}
\end{equation}
with
\begin{align}
	R_1^* &=  R_1\cos^2\left(\frac{\alpha}{2}\right)+ R_2\sin^2\left(\frac{\alpha}{2}\right)\\\nonumber
	M_{ss} &= \frac{M_0(1-E_1)\sin\alpha}{1-(E_1-E_2)\cos\alpha-E_1E_2}\\\nonumber
	&\underset{\text{TR}<<T_{1,2}}{=}~
	\frac{M_0\sin\alpha}{\left(\frac{R_1}{R_2}+1\right)-\cos\alpha\cdot\left(\frac{R_2}{R_1}-1\right)}\\\nonumber
	M_0^* &= M_o\sin\left(\frac{\alpha}{2}\right)
\end{align}
for
\begin{equation}
	E_{1/2}=e^{-R_{1/2}\cdot t}~.
\end{equation}
This can be reparameterized to the physical parameters $R_1$, $R_2$, $M_0$ and $\alpha$:
\begin{align}
	M(R_1, R_2, M_0, \alpha, t) &=
	\frac{M_0\sin\alpha}{\frac{R_2}{R_1}+1-\cos\alpha\left(\frac{R_2}{R_1}-1\right)}\\\nonumber
	&-\frac{M_0\sin\alpha\cdot e^{-R_1\cos^2\left(\frac{\alpha}{2}\right)\cdot t - R_2\sin^2\left(\frac{\alpha}{2}\right)\cdot t}}{\frac{R_2}{R_1}+1-\cos\alpha\left(\frac{R_2}{R_1}-1\right)}\\\nonumber
	&-M_0\sin\left(\frac{\alpha}{2}\right)\cdot e^{-R_1\cos^2\left(\frac{\alpha}{2}\right)\cdot t - R_2\sin^2\left(\frac{\alpha}{2}\right)\cdot t}
\end{align}
with its symbolic derivatives:
\begin{footnotesize}
	\begin{align}
		\frac{\partial M(R_1,R_2,M_0,\alpha,t)}{\partial R_1} &=
		-M_0R_2\sin\alpha\cdot(\cos\alpha-1)\cdot\frac{1-C(R_1, R_2, \alpha, t)}{B^2(R_1, R_2,\alpha)}\\\nonumber
		&~~+ M_0R_1 t \sin\alpha\cos\left(\frac{\alpha}{2}\right)\cdot\frac{C(R_1, R_2, \alpha, t)}{B(R_1, R_2,\alpha)}\\\nonumber
		&~~+ M_0 t \sin\left(\frac{\alpha}{2}\right)\cos^2\left(\frac{\alpha}{2}\right)\cdot C(R_1, R_2, \alpha, t)
	\end{align}
\end{footnotesize}
\begin{footnotesize}
	\begin{align}
		\frac{\partial M(R_1,R_2,M_0,\alpha,t)}{\partial R_2} &=
		M_0R_1\sin\alpha\cdot(\cos\alpha-1)\cdot\frac{1-C(R_1, R_2, \alpha, t)}{B^2(R_1, R_2,\alpha)}\\\nonumber
		&~~+ M_0R_1 t \sin\alpha\sin^2\left(\frac{\alpha}{2}\right)\cdot\frac{C(R_1, R_2, \alpha, t)}{B(R_1, R_2,\alpha)}\\\nonumber
		&~~+ M_0 t \sin^3\left(\frac{\alpha}{2}\right)\cdot C(R_1, R_2, \alpha, t)
	\end{align}
\end{footnotesize}
\begin{footnotesize}
	\begin{align}
		\frac{\partial M(R_1,R_2,M_0,\alpha,t)}{\partial \alpha} &=
		\frac{M_0R_1\sin^2\alpha(R_2-R_1)\cdot(C(R_1, R_2, \alpha, t)-1)}{B^2(R_1, R_2,\alpha)}\\\nonumber
		&~~- \frac{M_0R_1 \sin\alpha \cdot C(R_1, R_2, \alpha, t) t \sin\left(\frac{\alpha}{2}\right)\cos\left(\frac{\alpha}{2}\right)(R_1-R_2)}
		{B(R_1, R_2,\alpha)}\\\nonumber
		&~~- \frac{M_0 R_1\cos\alpha\cdot \left(C(R_1, R_2, \alpha, t)-1\right)}{B(R_1, R_2,\alpha)}\\\nonumber
		&~~- \frac{M_0}{2}\cos\left(\frac{\alpha}{2}\right)\cdot C(R_1, R_2, \alpha, t)\\\nonumber
		&~~- M_0\sin^2\left(\frac{\alpha}{2}\right)\cos\left(\frac{\alpha}{2}\right)(R_1-R_2)\cdot C(R_1, R_2, \alpha, t) t
	\end{align}
\end{footnotesize}
\begin{footnotesize}
	\begin{align}
		\frac{\partial M(R_1,R_2,\alpha,t)}{\partial M_0} &=
		\frac{\sin\alpha}{\frac{R_2}{R_1}+1-\cos\alpha\left(\frac{R_2}{R_1}-1\right)}\\\nonumber
		&-\frac{\sin\alpha\cdot e^{-R_1\cos^2\left(\frac{\alpha}{2}\right)\cdot t - R_2\sin^2\left(\frac{\alpha}{2}\right)\cdot t}}{\frac{R_2}{R_1}+1-\cos\alpha\left(\frac{R_2}{R_1}-1\right)}\\\nonumber
		&-\sin\left(\frac{\alpha}{2}\right)\cdot e^{-R_1\cos^2\left(\frac{\alpha}{2}\right)\cdot t - R_2\sin^2\left(\frac{\alpha}{2}\right)\cdot t}
	\end{align}
\end{footnotesize}
with 
\begin{align}
	B(R_1, R_2,\alpha) &= (R_1-R_2)\cos\alpha+R_1+R_2\\\nonumber
	C(R_1, R_2, \alpha, t) &= \exp\left(-R_1\cos^2\left(\frac{\alpha}{2}\right)\cdot t
	-R_2\sin^2\left(\frac{\alpha}{2}\right)\cdot t\right)~.
\end{align}

\newpage

\begin{sidewaystable}[!ht]
	\footnotesize
	\renewcommand{\arraystretch}{1.4}
	\caption{Table listing the sequence parameters for the performed measurements of this work.\label{tab::seq-parameters}}
	\begin{tabular*}{1.2\textheight}{@{\extracolsep\fill}c||cc|ccc|cc|cc@{\extracolsep\fill}}
		Sequence & \multicolumn{2}{c}{IR FLASH} & \multicolumn{3}{c}{IR bSSFP} & \multicolumn{2}{c}{Turbo FLASH} & \multicolumn{2}{c}{Spin-Echo} \\\hline
		Object & phantom & in vivo & phantom & in vivo & in vivo & phantom & in vivo & phantom & phantom \\\hline
		Figure & \ref{fig:irflash}.B & \ref{fig:irflash}.C, \ref{fig:irflash}.D, \ref{fig:irbssfp_invivo}.B & \ref{fig:irbssfp_nist} & \ref{fig:irbssfp_invivo} & \ref{fig:irbssfp_invivo} & \ref{fig:irbssfp_nist} & \ref{fig:irbssfp_invivo} & \ref{fig:irbssfp_nist} & \ref{fig:irbssfp_nist}\\\hline
		TR$|$TE [ms] & 4.1$|$1.84 & 4.1$|$2.58 & 4.88$|$2.44 & 4.88$|$2.44 & 10.8$|$5.4 & 2000$|$2.14 & 2000$|$2.14 & 8000$|$15 & 8000$|$(15:40:455) \\\hline
		FA [$^{\circ}$] & 6 & 6 & 45 & 45 & 45 & 8 & 8 & - & - \\\hline
		$T_{\text{RF}}$ [ms] & 1 & 1 & 1 & 1 & 2.5 & - & - & - & - \\\hline
		Nominal Slice Thickness [mm] & 5 & 5 & 5 & 5 & 5 & 5 & 5 & 5 & 5 \\\hline
		Repetitions & 1020 & 1000 & 1000 & 1000 & 500 & 1 & 1 & 1 & 1\\\hline
		Coils & 20 & 20 & 20 & 20 & 20 & 20 & 20 & 20 & 20\\\hline
		BWTP & 4 & 4 & 4 & 4 & 4 & - & - & - & - \\\hline
		BR & 256 & 256 & 192 & 256 & 256 & 192 & 256 & 256 & 256 \\\hline
		FoV [mm] & 192 & 200 & 200 & 200 & 200 & 200 & 200 & 200 & 200 \\\hline
		Duration [min:s] & 0:04 & 0.04 & 0:05 & 0:05 & 0:06 & 0:04 & 0:04 & 34:16 & 34:16\\\hline
		others & \#tiny GA=7 & \#tiny GA=7 & \#tiny GA=7 & \#tiny GA=7 & \#tiny GA=7 & & & $T_{\text{inv}}$=30:250:2530 ms & \\
	\end{tabular*}
\end{sidewaystable}

\newpage

\begin{figure}[t]
	\centerline{\includegraphics[width=\textwidth]{forward_operator.pdf}}
	\caption{Illustration of the operators used in the Bloch model-based reconstruction.
		The bottom part presents the ODEs for the signal and the derivatives.}
	\label{fig::overview}
\end{figure}

\newpage

\begin{figure}[t]
	\centerline{\includegraphics[width=\textwidth]{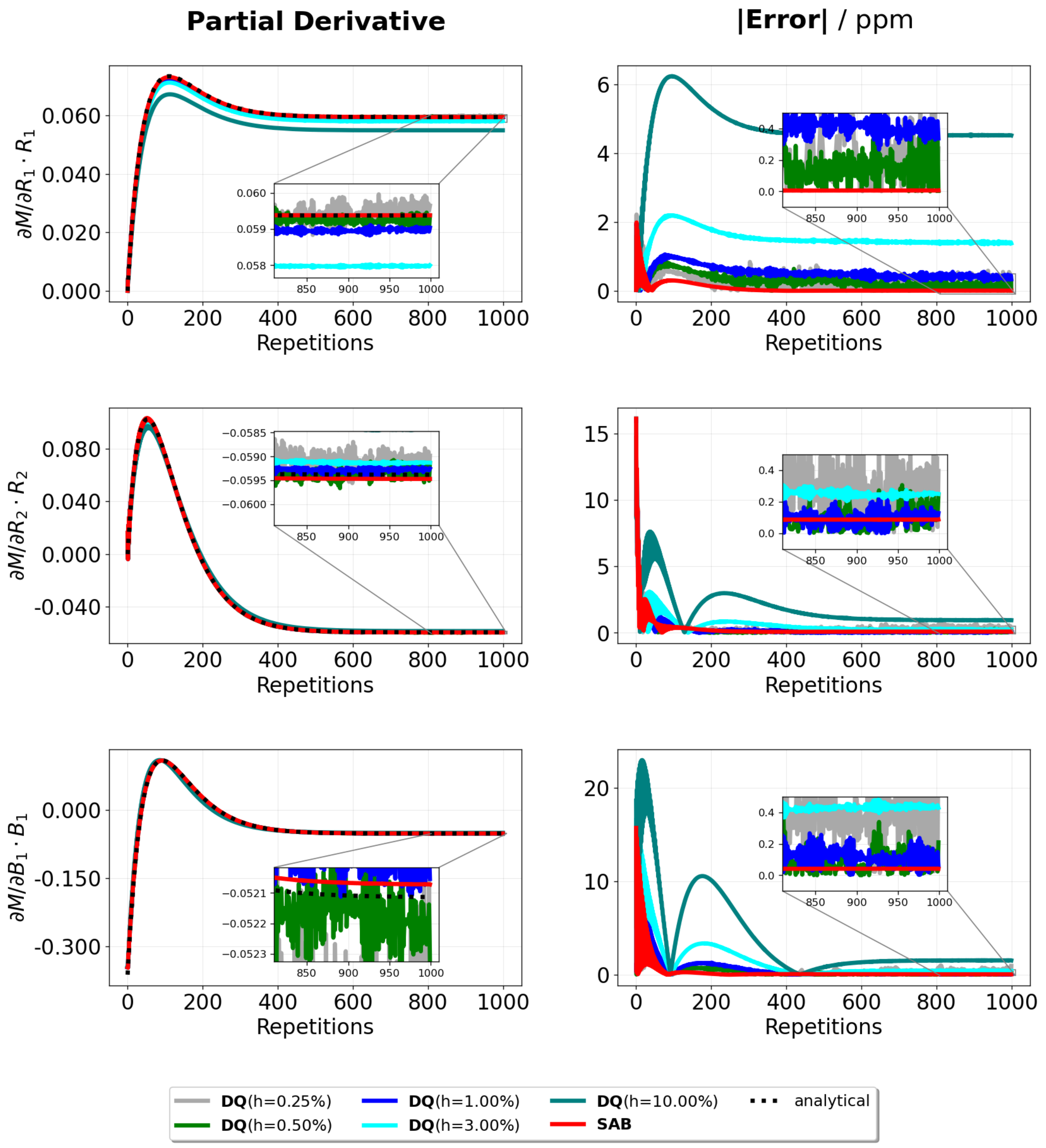}}
	\caption{\textbf{Left}: Temporal evolution of the partial derivatives
		with respect to $R_1$, $R_2$ and $B_1$ estimated for an
		IR bSSFP sequence with the SAB, DQ with varying perturbation
		levels $h$ and the analytical references.
		\textbf{Right}: Plot with point-wise errors of the various DQ methods and the
		SAB with respect to the analytical reference.
		Note that the errors are presented in ppm for visualization.}
	\label{fig::sab}
\end{figure}

\newpage

\begin{figure}[t]
	\centerline{\includegraphics[width=\textwidth]{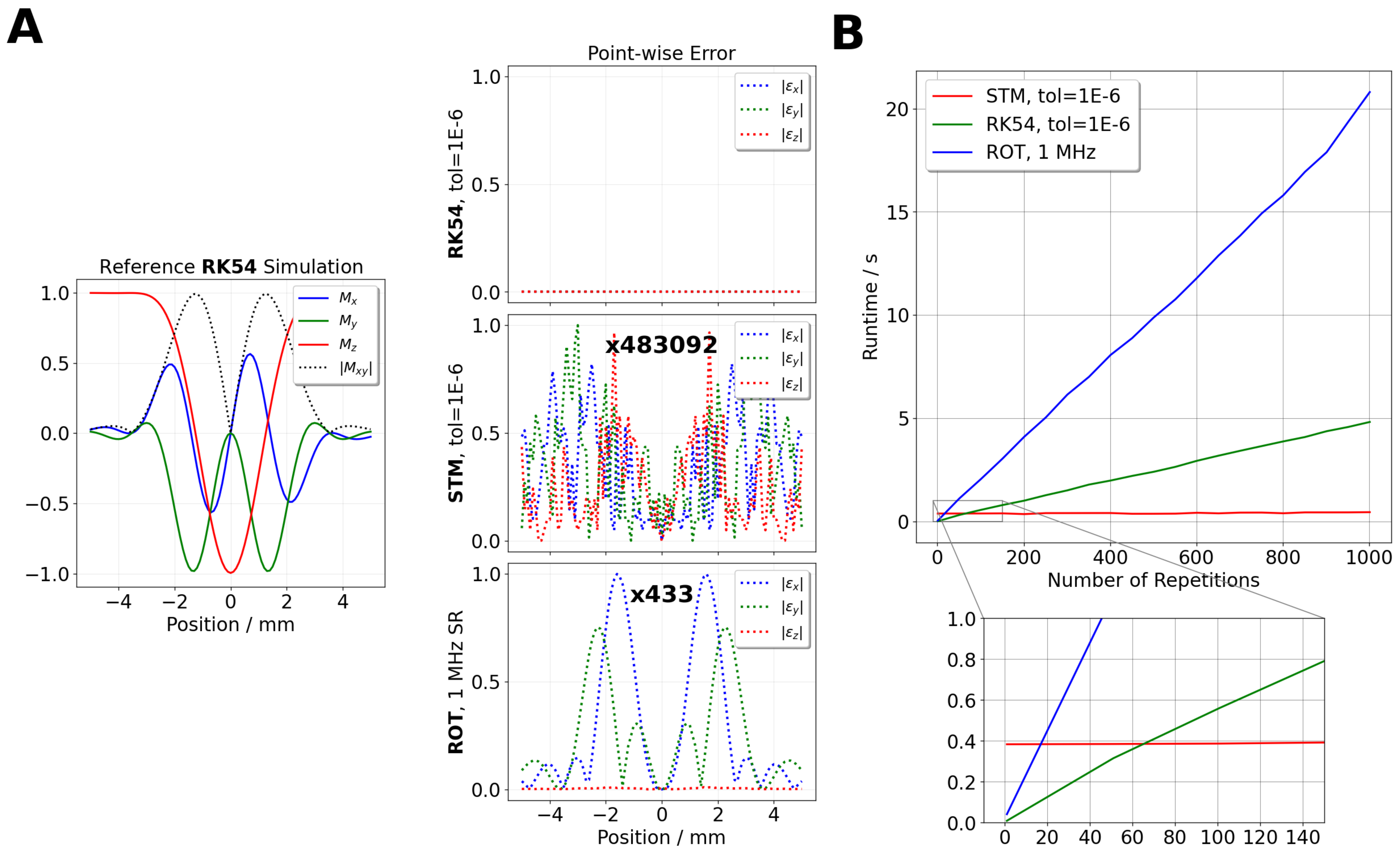}}
	\caption{\textbf{A}: The slice-selection gradient based simulation
		for a Hamming-windowed sinc-shaped
		inversion pulse simulated with the RK54 framework is shown (left).
		The point-wise errors of the RK54 (top), the STM technique (center) and the ROT method (bottom) with
		sampling rate 1 MHz are plotted for the x-, y- and
		z- component of the magnetization.
		Note that the errors are scaled by large factors for visualization.
		\textbf{B}: The runtime of the STM technique is compared to the
		reference RK54 method and a ROT simulation performed with a sampling rate of 1 MHz.
		The simulation is performed for 101 isochromats
		homogeneously distributed along a slice-selection gradient during a
		FLASH sequence for various numbers of repetitions.
		The maximum of 1000 is chosen to cover about 4 s of acquisition, required to
		measure enough data points for mapping high $T_1$ values.
		A more detailed version of this figure has been added to the Supplementary Section S3.}
	\label{fig::stm}
\end{figure}

\newpage

\begin{figure}[t]
	\centering
	\includegraphics[width=.75\textwidth]{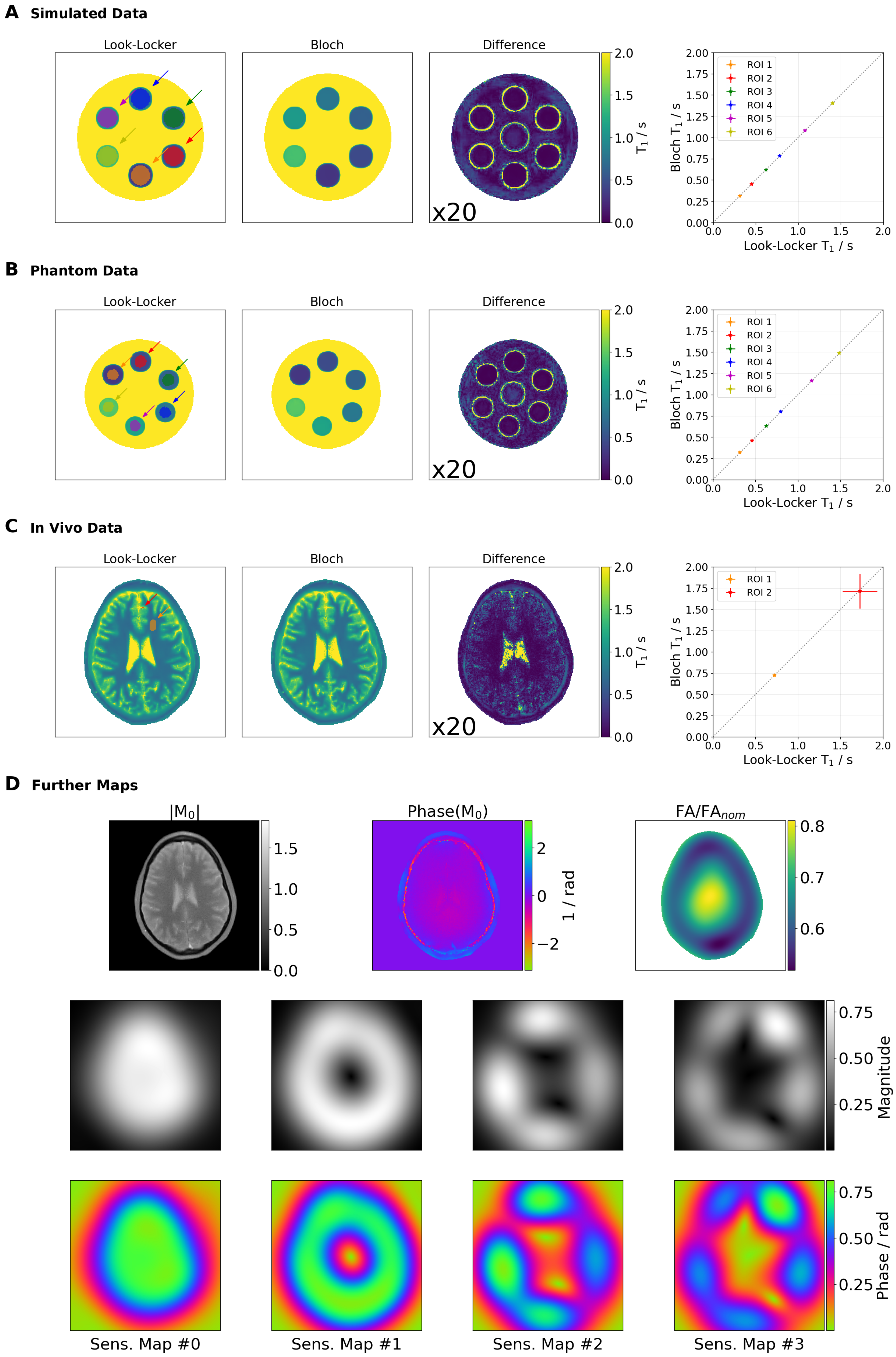}
	\caption{Reconstructed $T_1$ parameter maps for radial single-shot IR FLASH data
		acquired from a numerical (\textbf{A}) and measured phantom (\textbf{B}) as well as a human brain (\textbf{C}).
		The Bloch model-based reconstruction and the differences between the
		two methods are shown in the middle. The difference map is scaled up 
		by a factor of 20 to improve visualization.
		On the right the $T_1$ values of the color-coded ROIs (arrows) 
		of the Bloch reconstruction vs. Look-Locker reconstruction
		are plotted together with standard deviations.
		Besides the in vivo $T_1$ map presented in \textbf{C}, the Bloch model-based technique
		reconstructs a complex valued $M_0$ map, a relative flip angle map and
		complex coil sensitivities shown in \textbf{D}. }
	\label{fig:irflash}
\end{figure}

\newpage

\begin{figure}[t]
	\centering
	\includegraphics[width=\textwidth]{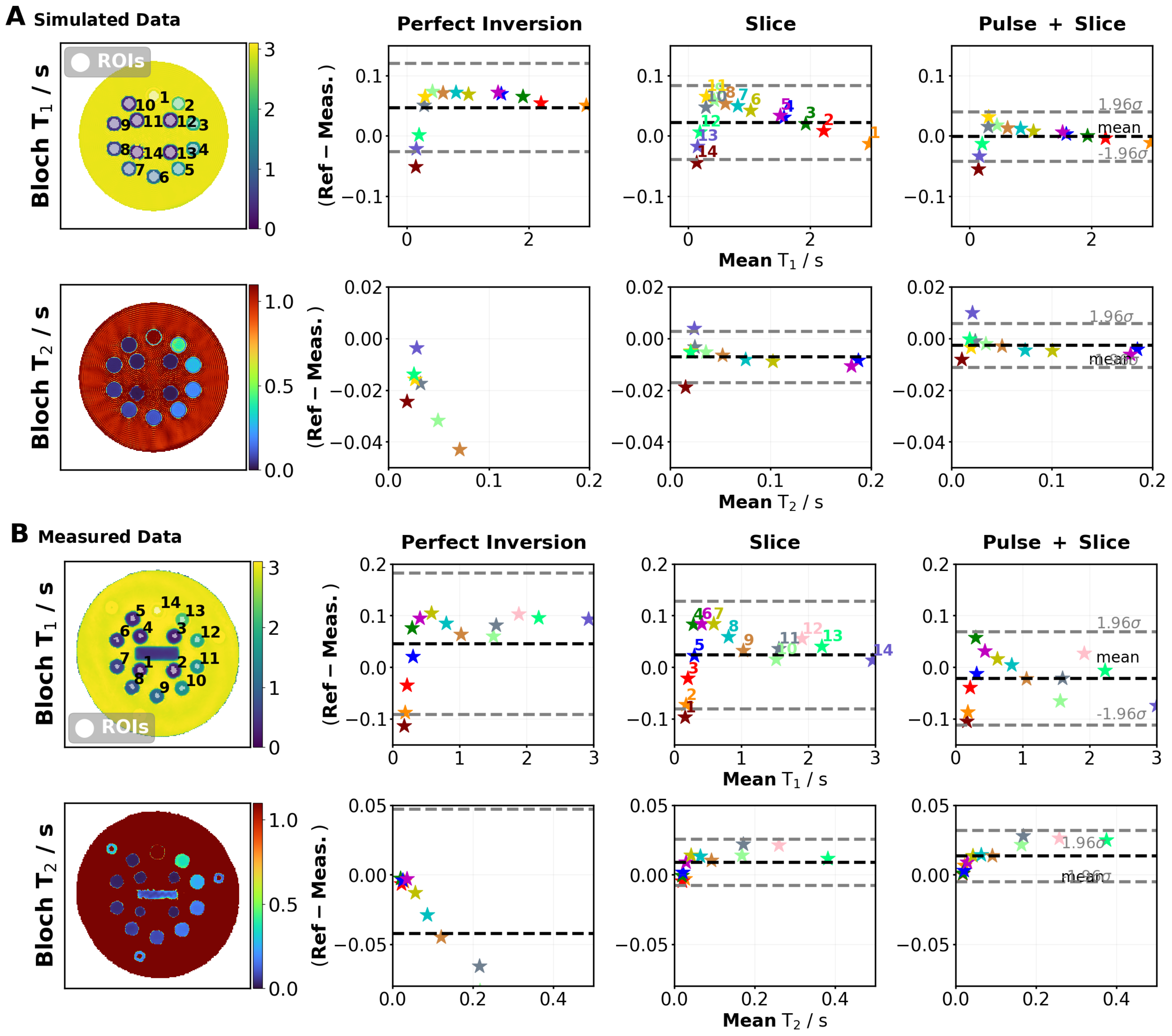}
	\caption{ \textbf{A}: Reconstructed $T_1$ and $T_2$ parameter maps and the corresponding
		ROI values for numerical radial single-shot IR bSSFP data
		of a digital multi-coil reference object simulated in k-space.
		The left side shows the reconstructed parameter maps and the right the ROI analysis
		results in Bland-Altman plots relative to the simulated reference values.
		The analyzed ROIs are marked and numbered in the $T_1$ map.
		\textbf{B}: Reconstructed $T_1$ and $T_2$ parameter maps and the corresponding
		ROI values for a radial single-shot IR bSSFP measurement of the $T_2$ spheres
		of the NIST system phantom.
		The left side corresponds to the rightmost ROI analysis.
		The right side presents the comparison of the analyzed ROIs
		reconstructed with the Bloch model-based technique
		for various signal model assumptions
		compared with gold-standard reference values in Bland-Altman plots.
		For improved visualization individual outliers are removed from the plot.}
	\label{fig:irbssfp_nist}
\end{figure}

\newpage

\begin{figure}[t]
	\centerline{\includegraphics[width=\textwidth]{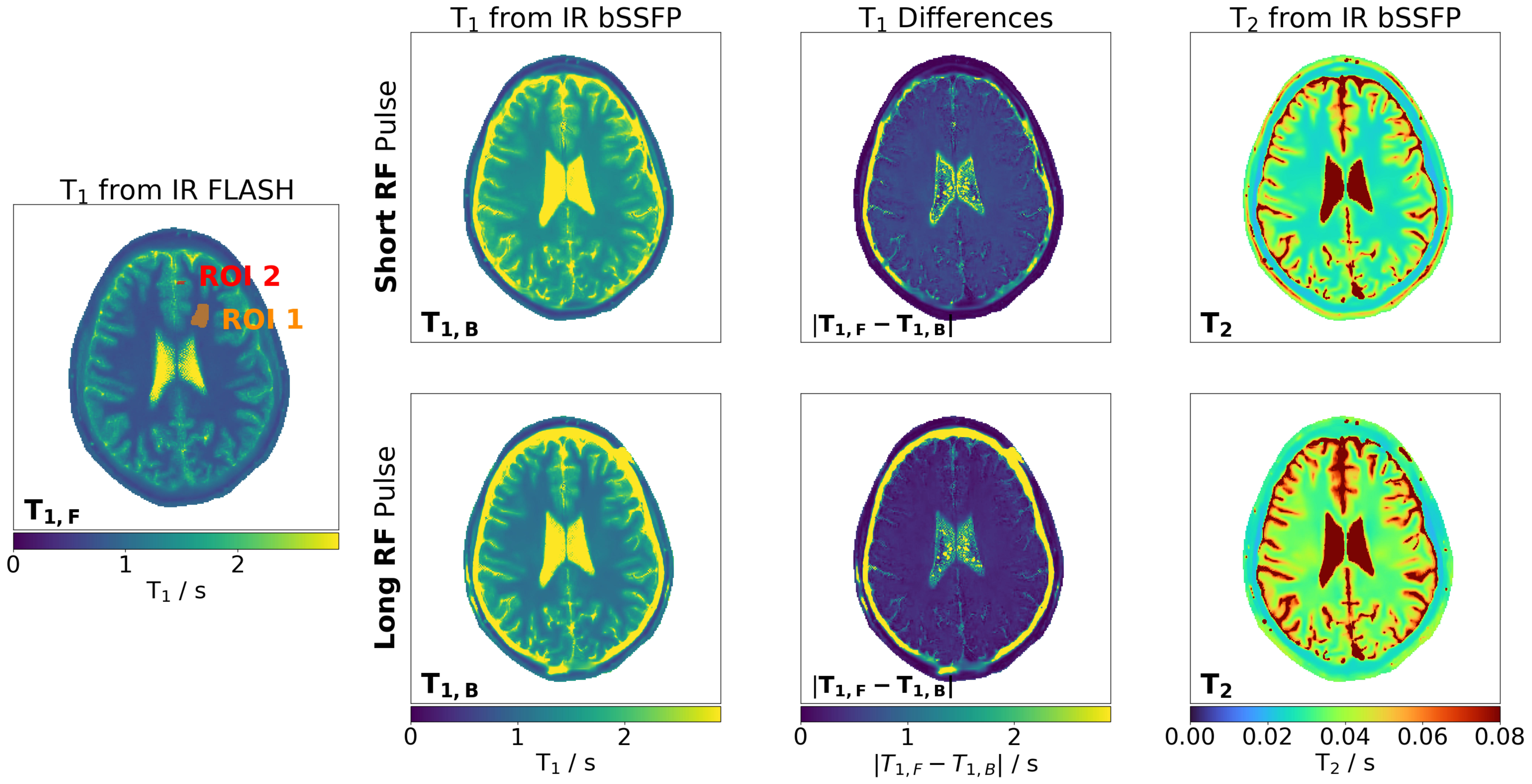}}
	\caption{The $T_1$ parameter map reconstructed from a radial
		single-shot IR FLASH in vivo dataset with a Look-Locker model-based reconstruction
		is shown on the left.
		It also shows the reconstructed $T_1$ parameter maps of a radial
		single-shot IR bSSFP in vivo dataset acquired on the same brain slice
		for short RF pulses ($D_{\text{RF}}$: 1 ms, TR: 4.88 ms) on the top and
		long RF pulses ($D_{\text{RF}}$: 2.5 ms, TR: 10.8 ms) on the bottom
		reconstructed with the Bloch model-based reconstruction.
		In the center column the difference maps are shown.
		The values corresponding to the colored ROIs are listed in Table \ref{tab::invivo-irbssfp}.
		On the right the $T_2$ parameter maps for the short and long RF pulse experiments
		are shown reconstructed from the IR bSSFP sequence.}
	\label{fig:irbssfp_invivo}
\end{figure}

\newpage

\begin{table}[!ht]
	\renewcommand{\arraystretch}{1.4}
	\centering
	\caption{Table listing the single-shot in-vivo IR bSSFP ROI analysis results presented in Figure \ref{fig:irbssfp_invivo}.
		\label{tab::invivo-irbssfp}}
	\makebox[\textwidth][c]{%
	\begin{tabular}{c|c|c|c|c|c} & $T_{1,\text{IR FLASH}}$ [s] & $T_{1,\text{IR bSSFP, short}}$ [s] & $T_{1,\text{IR FLASH, long}}$ [s] & $T_{2,\text{IR bSSFP, short}}$ [s] & $T_{2,\text{IR bSSFP, long}}$ [s] \\\hline
		\textbf{ROI 1} & 0.737$\pm$0.016&1.35$\pm$0.019&1.061$\pm$0.019&0.024$\pm$0.001&0.033$\pm$0.001\\
		\textbf{ROI 2} & 1.736$\pm$0.299&2.434$\pm$0.413&2.226$\pm$0.456&0.066$\pm$0.02&0.096$\pm$0.088
	\end{tabular}
	}
\end{table}

\FloatBarrier
\newpage

\section*{Supporting Information}

\renewcommand{\thetable}{S\arabic{table}}
\renewcommand{\thefigure}{S\arabic{figure}}
\renewcommand{\thesection}{S\arabic{section}}
\setcounter{section}{0}
\setcounter{figure}{0}
\setcounter{table}{0}

\section{Supporting Figure S1}

\begin{figure}[!ht]
	\centering
	\includegraphics[width=.9\textwidth]{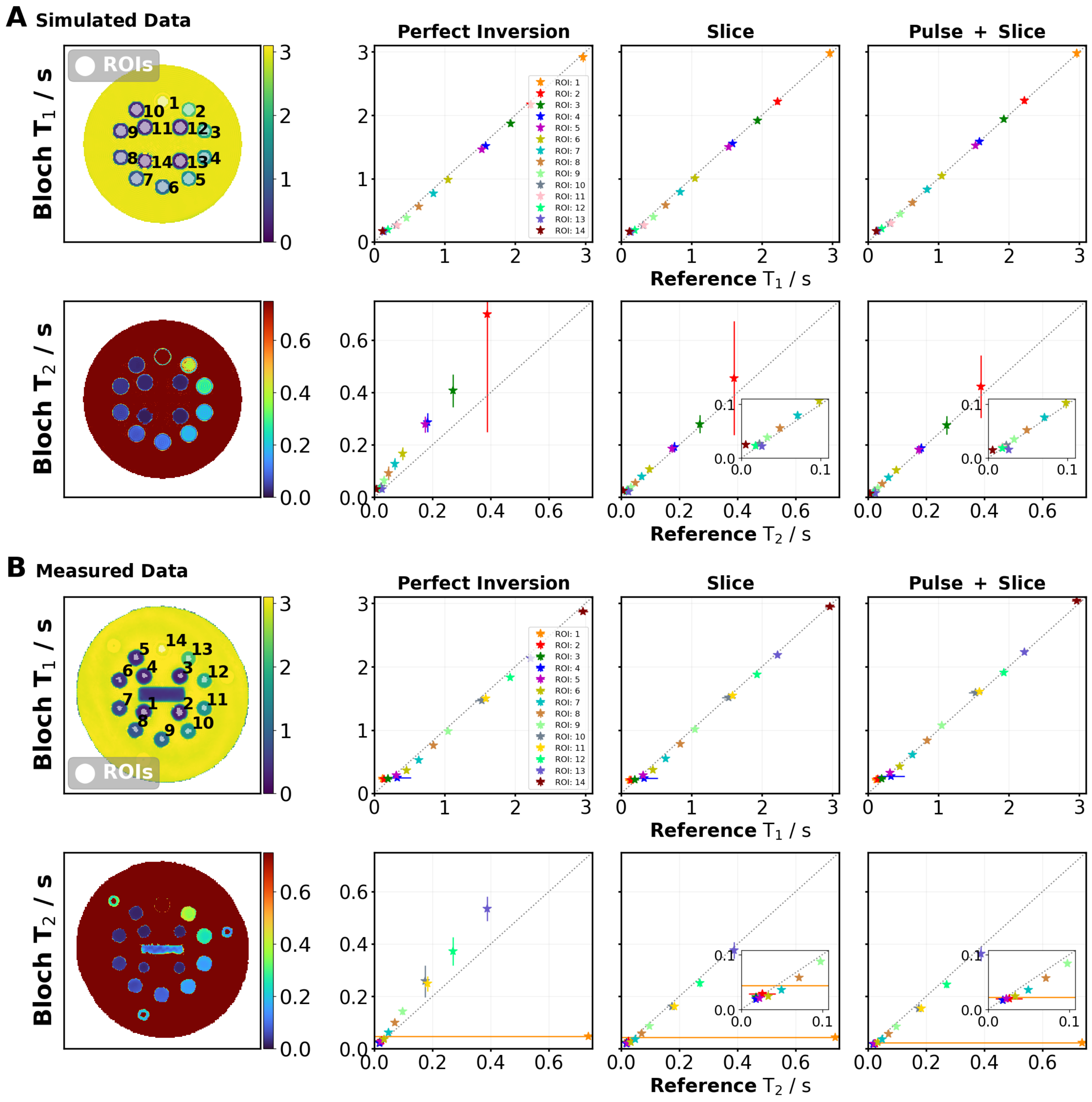}
	\caption{ \textbf{A}: Reconstructed $T_1$ and $T_2$ parameter maps and the corresponding
		ROI values for numerical radial single-shot IR bSSFP data
		of a digital multi-coil reference object simulated in k-space.
		The left side shows the reconstructed parameter maps and the right the ROI analysis
		results relative to the simulated reference values wit standard derivations.
		The analyzed ROIs are marked and numbered in the $T_1$ map.
		\textbf{B}: Reconstructed $T_1$ and $T_2$ parameter maps and the corresponding
		ROI values for a radial single-shot IR bSSFP measurement of the $T_2$ spheres
		of the NIST system phantom.
		The left side corresponds to the rightmost ROI analysis.
		The right side presents the comparison of the analyzed ROIs
		reconstructed with the Bloch model-based technique
		for various signal model assumptions
		compared with gold-standard reference values.}
	\label{fig:irbssfp_nist_diag}
\end{figure}

\FloatBarrier
\newpage
\section{Supporting Figure S2}

\begin{figure}[!ht]
	\centering
	\includegraphics[width=\textwidth]{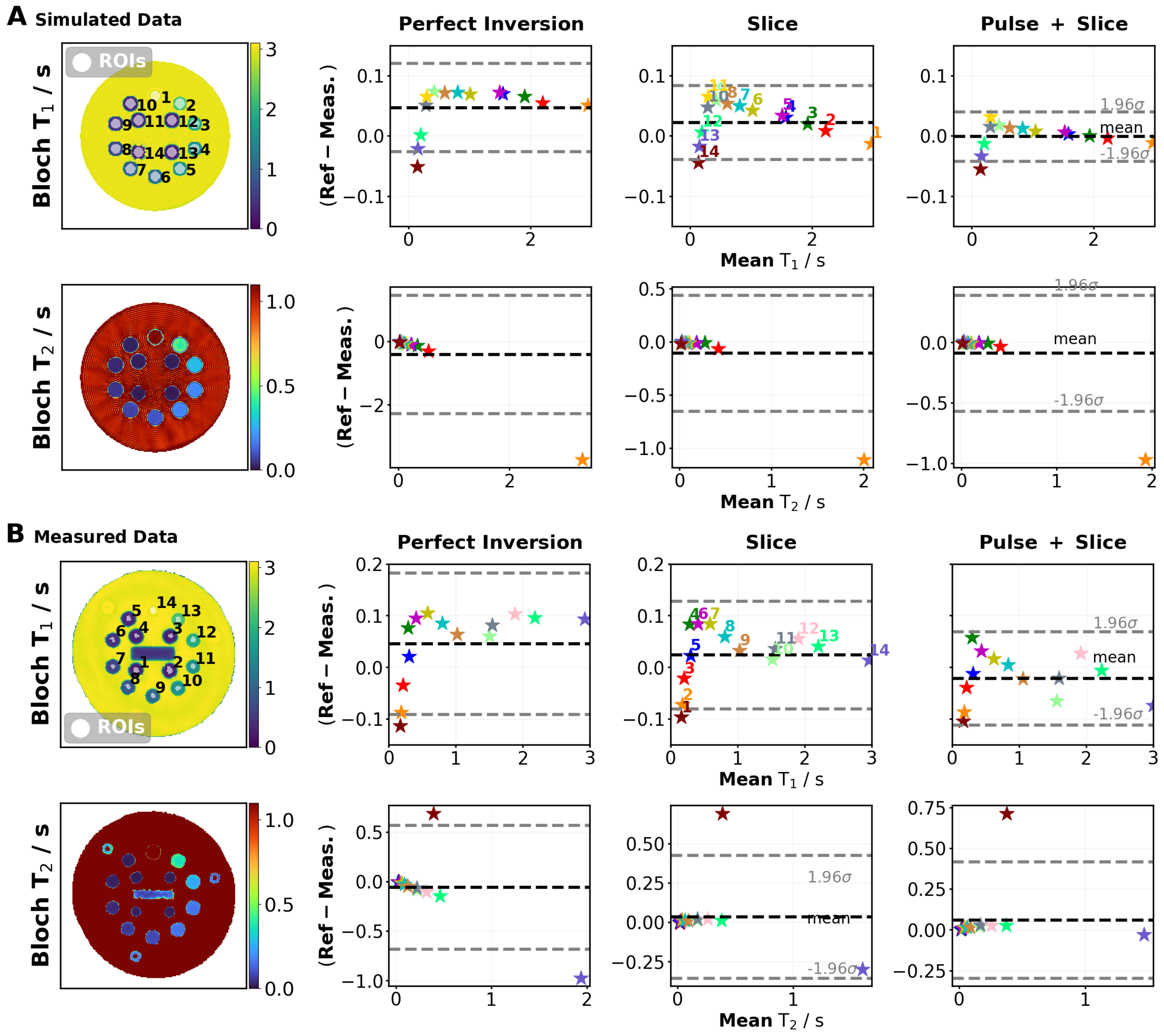}
	\caption{\textbf{A}: Reconstructed $T_1$ and $T_2$ parameter maps and all corresponding
		ROI values for simulated radial single-shot IR bSSFP data.
		Here, the datapoints removed to improve visualization in Figure 5 
		are included. Reconstructed parameter maps are shown on the left
		Bland-Altman plots using the simulated reference values are shown on the right.
		The ROIs are marked and numbered in the $T_1$ map.
		\textbf{B}: Reconstructed $T_1$ and $T_2$ parameter maps and the corresponding
		ROI values for a radial single-shot IR bSSFP measurement of the $T_2$ spheres
		of the NIST system phantom.
		The left side corresponds to the rightmost ROI analysis.
		The right side shows the Bland-Altman plots comparing the
		values of the ROIs reconstructed with the Bloch model-based technique
		for various assumptions used in the signal model
		compared with gold-standard reference values.
		For improved visualization individual outliers were removed from the plot.}
	\label{fig:irbssfp_nist_bland_all}
\end{figure}

\FloatBarrier
\newpage
\section{Supporting Figure S3: Simulation Accuracy}
\label{sec::detailed_sim_acc}

\begin{figure}[!ht]
	\centering
	\includegraphics[width=\textwidth]{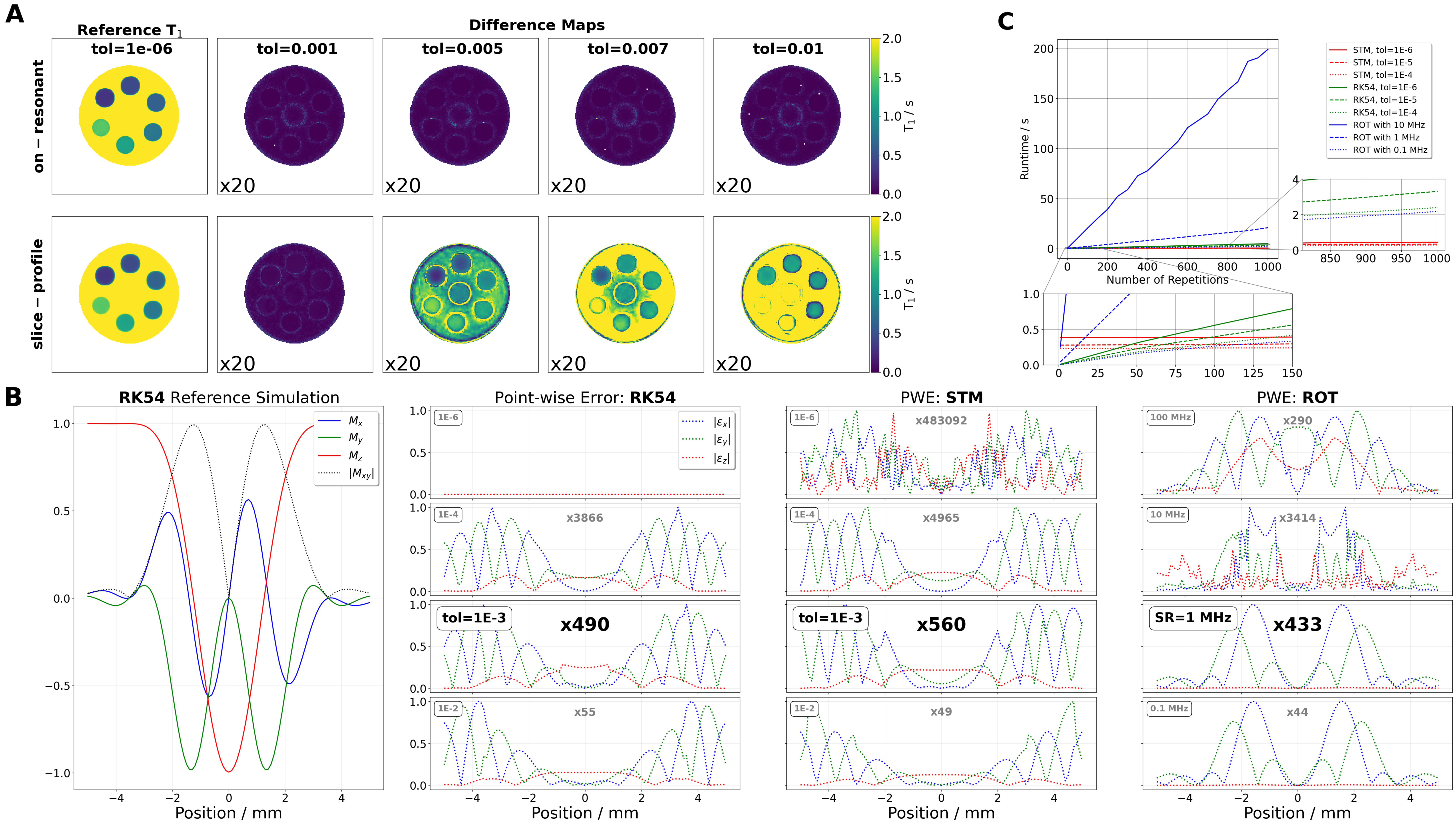}
	\caption{\textbf{A}: $T_1$ maps reconstructed from
		single-shot IR FLASH (same as for Figure 4.B) 
		for varying
		tolerance for computation of STMs. The forward model assumes on-resonant spins.
		The difference of the reconstructions for tolerances of 1e-5 to 1e-2 compared
		to the 1e-6 reference are shown in the lower row. The differences are scaled
		by large factors for improved visualization.
		\textbf{B}: Same analysis as for part A, but assuming a complex forward model
		including a slice-selection gradient.
		\textbf{C}: Extension of Figure 3.A 
		with varying tolerances
		for RK54 and STM and sampling rates for ROT.
		\textbf{D}: Extension of Figure 3.B 
		with varying tolerances
		for RK54 and STM as sampling rates for ROT.
	}
	\label{fig:simulation_accuracy}
\end{figure}

This section discusses the simulation accuracy of the ODE solvers
for different tolerances and sampling rates in more detail.
In Figure \ref{fig:simulation_accuracy}.A reconstructions of
the IR FLASH phantom from Figure 4.B 
using the Bloch model-based reconstruction with signal model
assumptions of only on-resonant spins and with spins distributed along a
slice-selection gradient are shown.
The reconstructions are performed with varying tolerance
for the STM simulation method while keeping the initial step-size
of the RK54 solver constant at 1E-4.

For the on-resonance results the difference maps are small.
Here, the reconstructed parameter maps for higher error
tolerances are very similar to the reference map.
For the signal model with simulation of the slice-selection gradient
the reconstructed parameter maps are similar up to a
tolerance of 0.001, while lower values result in large variations in the $T_1$ parameter map.
The important result is,
that the error of the simulation from an STM tolerance
value of 0.001 still allows for accurate reconstructions even for
complex spin dynamics.

In Figure \ref{fig:simulation_accuracy}.B the error plot from Figure 3 
is extend to include more values for tolerance and sampling rate.
To translate the required STM tolerance of 0.001 from Figure \ref{fig:simulation_accuracy}.A
for complex dynamics to the required sampling rate
for the ROT simulation, it demonstrates the point-wise errors (PWE) for RK54,
STM and ROT for the same simulation with slice-selection gradient.
The tolerance value for the estimated limit of STM of 0.001 produces 
an error similar to the 1 MHz sampling rate.
This leads to the conclusion that at least a sampling rate of 1 MHz
is required for accurate reconstruction of complex spin dynamics
involving slice-selection gradients.

The simulation times for the same analysis with slice-selection gradient
as presented in Figure 3 
are shown in \ref{fig:simulation_accuracy}.C
for multiple solver tolerances and sampling rates.

\FloatBarrier
\newpage
\section{Supporting Figure S4}

\begin{figure}[!ht]
	\centering
	\includegraphics[width=\textwidth]{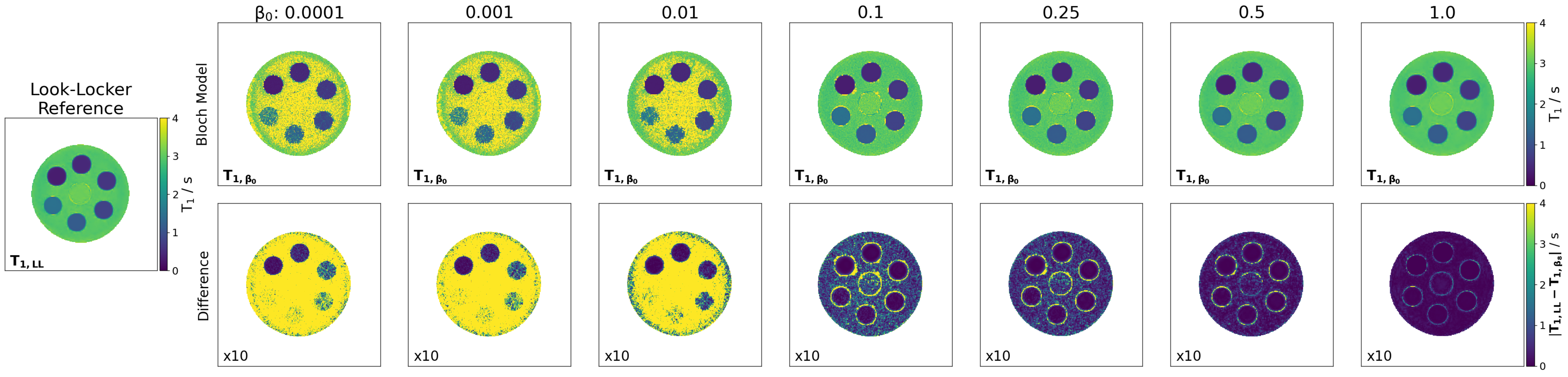}
	\caption{Reconstructed $T_1$ parameter maps similar to Figure 4.B 
		for varying
		wavelet regularization strengths $\beta_0$ for the Bloch model-based reconstruction (upper row).
		On the left a reference reconstruction estimated with
		the re-parameterized Look-Locker model-based technique is shown.
		In the bottom row the differences between Bloch model-based reconstruction
		and the reference map are shown scaled by a factor of 10 for improved
		visualization.
	}
	\label{fig:wav_reg_test}
\end{figure}

\FloatBarrier
\newpage
\section{Supporting Figure S5: Influence of the Magnetization Transfer Effect}
\label{sec:mt_influence}

To get an estimate of the magnetization transfer effect (MT) influence on the reconstructed $T_1$ parameter
maps, multiple single-shot IR bSSFP experiments were performed with varying TR and $T_{RF}$.
The sequence parameters are shown in Table \ref{tab::mt_fit_seq}.
The analytical representation of the signal behaviour during an IR bSSFP experiment
can be described by the equation F10. 
Following \cite{Ehses_Magn.Reson.Med._2013} the MT effect reduces the $T_1^*$ as well as the
steady-state magnetization $M_{ss}$, while not affecting $M_0$.
They propose an exponential model for correction of the effect of MT on $M_{ss}$:

\begin{equation}
	M_{{ss}}(\beta) = \left(M^{\text{no MT}}_{{ss}}-M^{\text{full MT}}_{{ss}}\right)
	\cdot\left(1-e^{-k\cdot\beta}\right) + M^{\text{full MT}}_{{ss}}
\end{equation}	

Here, the signal which is not affected by MT is denoted as $M^{\text{no MT}}_{ss}$,
the signal from a fully-saturated solid pool is $M^{\text{full MT}}_{ss}$,
the rate constant is $k$ and the increase in RF pulse duration is $\beta$.

The $T_1$ value for an IR bSSFP can be estimated by

\begin{equation}
	T_1 = \frac{T_1^*}{M_{{ss}}}M_0\cos(\frac{\alpha}{2})
\end{equation}

with the flip angle $\alpha$.

Taking into account that $T_1^*$ is less affected by MT than $M_{ss}$
\cite{Ehses_Magn.Reson.Med._2013}, we assume that the MT influence on $T_1$
follows the model: 

\begin{equation}
	T_{1}(\beta) \approx \frac{1}{\left(a-b\right)
		\cdot\left(1-e^{-k'\cdot\beta}\right) + b}~.
	\label{eq:t1_exp_model}
\end{equation}

\begin{figure}[!ht]
	\centering
	\includegraphics[width=.7\textwidth]{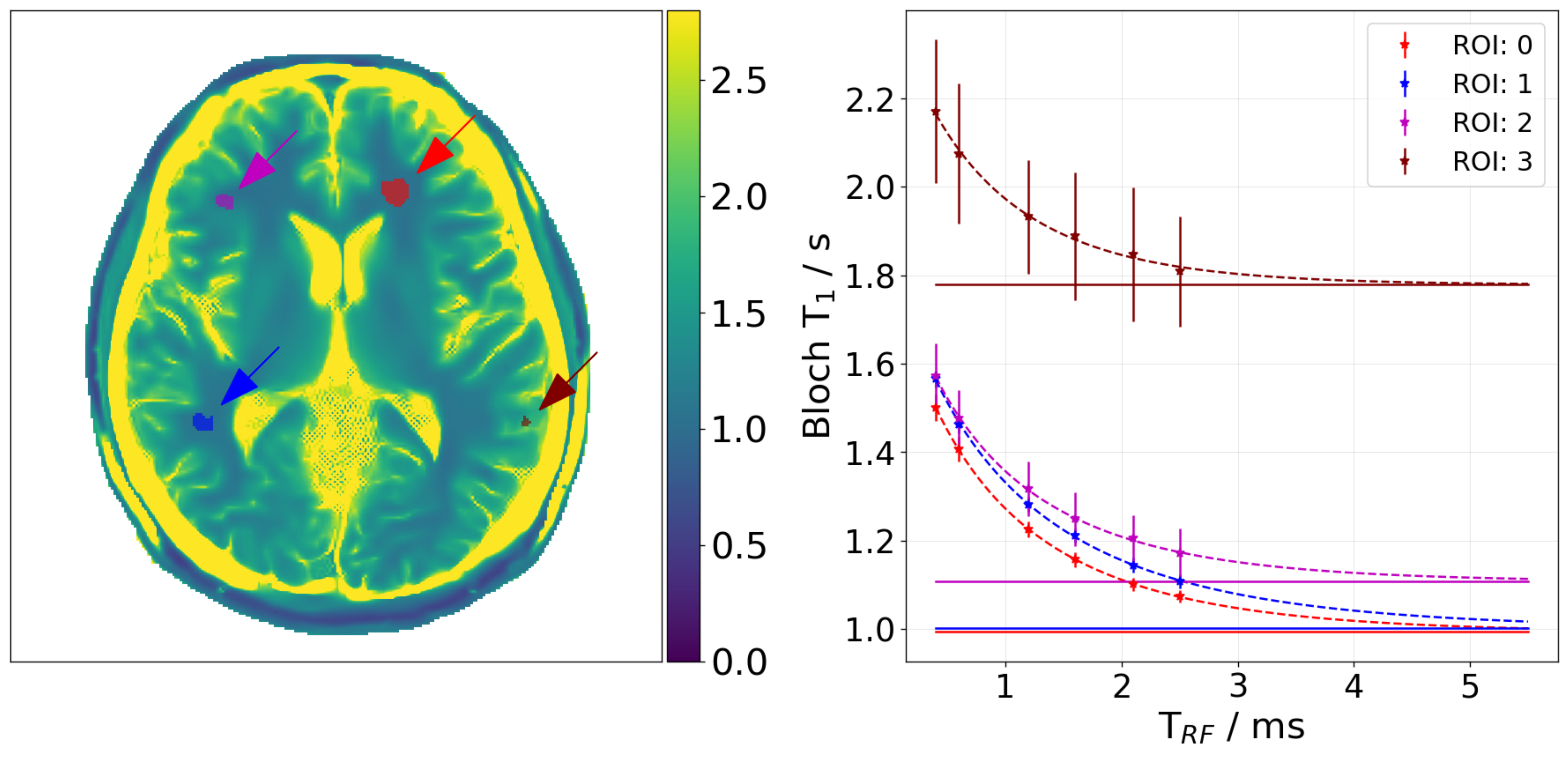}
	\caption{Visualization of the magnetization transfer effect in $T_1$ maps
		reconstructed from an IR bSSFP sequence with varying TR and $T_{RF}$.
		On the left the $T_1$ map corresponding to the longest $T_{RF}=2.5$ ms is plotted with colored ROIs. The mean value of the relaxation parameter values for the colored areas is plotted with its standard deviation on the right for various length of RF-pulses $T_{RF}$. The dotted lines represent the fitted values following the analysis of section \ref{sec:mt_influence}.
	}
	\label{fig:multi_trf_analysis}
\end{figure}

Equation \ref{eq:t1_exp_model} is fitted to the different relaxation parameters
from various acquisitions with different TR and $T_{RF}$ (Figure \ref{fig:multi_trf_analysis}).

The estimated parameters $a$, $b$ and $k'$ have no direct physical
interpretation, but can be used to interpolate an MT-free $T_1$ for $\beta\rightarrow\infty$.
The parameters of the fit are shown in Table \ref{tab::mt_fit_para}.

The results for $T_{1}(\beta\rightarrow\infty)$ suggest that the MT effect plays a significant role in
the $T_1$ offset observed in Figure 6 
, but also indicates that it
might not explain the complete discrepancy.

\FloatBarrier
\clearpage
\section{Supporting Figure S6}

\begin{figure}[!ht]
	\centering
	\includegraphics[width=.6\textwidth]{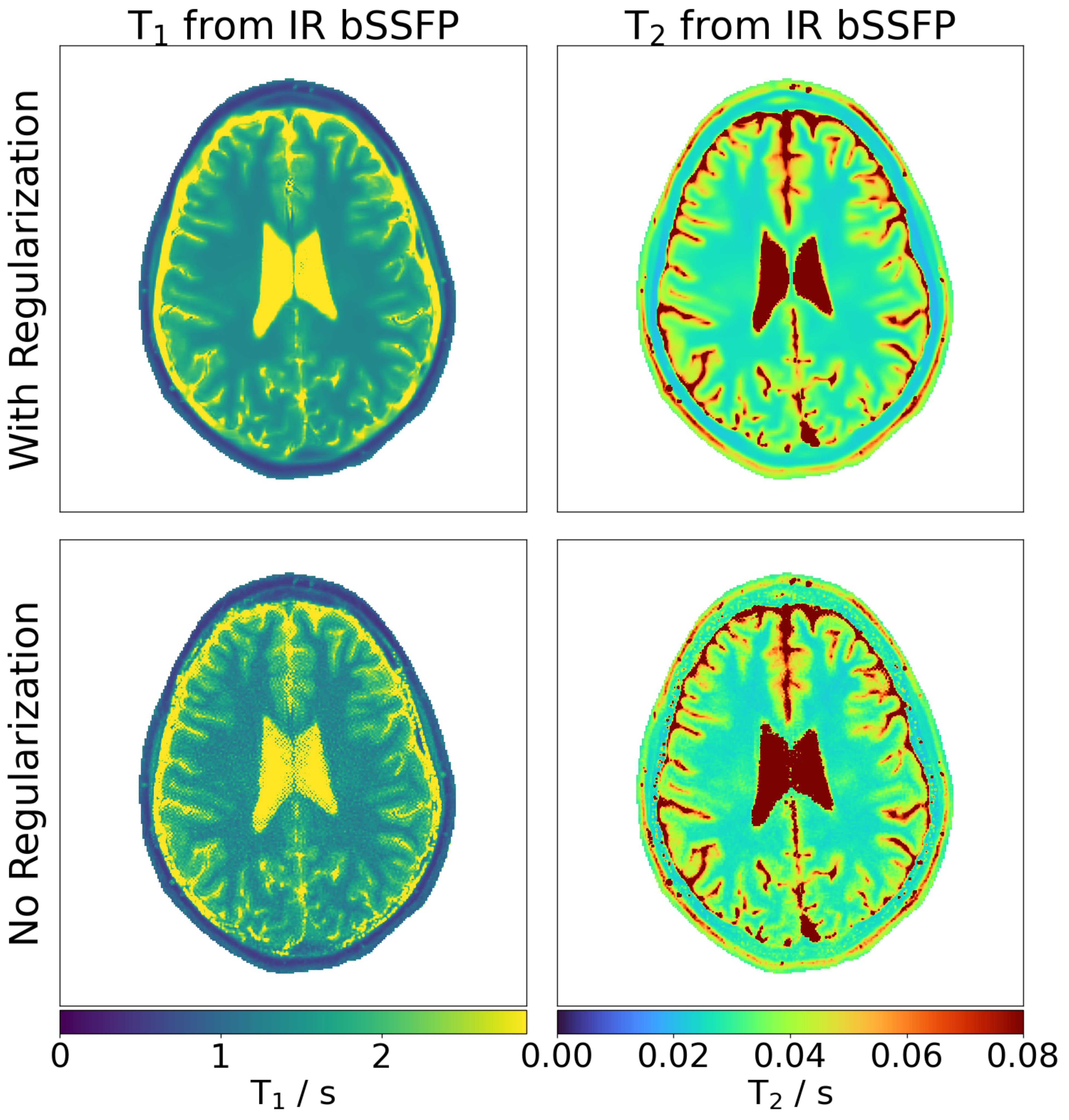}
	\caption{Bloch model-based reconstruction of the short RF pulse
		in vivo single-shot IR bSSFP dataset. The coil-sensitivities have been
		estimated with the regular method and have been added as fixed prior knowledge
		to both reconstruction. The top presents the regularized reconstruction,
		while the bottom results do not include any regularization on the
		parameter maps.}
	\label{fig:input_coils}
\end{figure}

\FloatBarrier
\newpage
\section{Supporting Table S1}

\begin{table}[!ht]
	\renewcommand{\arraystretch}{1.4}
	\centering
	\caption{Table listing the fitting parameters estimated for Figure \ref{fig:multi_trf_analysis}.
		\label{tab::mt_fit_para}}
	\begin{tabular}{c|c|c|c} & $a$ [s] & $b$ [s] & $k'$ [s] \\\hline
		\textbf{ROI 1} & 0.931$\pm$0.006&0.510$\pm$0.003&0.723$\pm$0.023\\
		\textbf{ROI 2} & 0.923$\pm$0.013&0.498$\pm$0.005&0.620$\pm$0.041\\
		\textbf{ROI 3} & 0.834$\pm$0.01&0.498$\pm$0.006&0.786$\pm$0.059\\
		\textbf{ROI 4} & 0.519$\pm$0.007&0.382$\pm$0.008&0.992$\pm$0.158\\
	\end{tabular}
\end{table}

\FloatBarrier
\newpage
\section{Supporting Table S2}

\begin{sidewaystable}[t]
	\renewcommand{\arraystretch}{1.4}
	\centering
	\caption{Table listing the sequence parameters for the analysis in Figure \ref{fig:multi_trf_analysis}.
		\label{tab::mt_fit_seq}}
	\begin{tabular*}{\textheight}{@{\extracolsep\fill}c||cccccc@{\extracolsep\fill}}
		Sequence & IR bSSFP & IR bSSFP & IR bSSFP & IR bSSFP & IR bSSFP & IR bSSFP\\\hline
		Object & in vivo & in vivo & in vivo & in vivo & in vivo & in vivo\\\hline
		TR$|$TE [ms] & 3.8$|$1.9 & 4.0$|$2.0 & 4.6$|$2.3 & 5$|$2.5 & 5.5$|$2.75 & 6.14$|$3.07 \\\hline
		FA [$^{\circ}$] & 35 & 35 & 35 & 35 & 35 & 35 \\\hline
		$T_{\text{RF}}$ [ms] & 0.4 & 0.6 & 1.2 & 1.6 & 2.1 & 2.5 \\\hline
		Nominal Slice Thickness [mm] & 5 & 5 & 5 & 5 & 5 & 5 \\\hline
		Repetitions & 1000 & 1000 & 1000 & 1000 & 1000 & 1000 \\\hline
		Coils & 20 & 20 & 20 & 20 & 20 & 20 \\\hline
		BWTP & 1 & 1 & 1 & 1 & 1 & 1 \\\hline
		BR & 256 & 256 & 256 & 256 & 256 & 256 \\\hline
		FoV [mm] & 200 & 200 & 200 & 200 & 200 & 200 \\\hline
		Duration [min:s] & 0:04 & 0.04 & 0:05 & 0:05 & 0:06 & 0:06 \\\hline
		others & \#tiny GA=13 & \#tiny GA=13 & \#tiny GA=13 & \#tiny GA=13 & \#tiny GA=14 & \#tiny GA=13 \\
	\end{tabular*}
\end{sidewaystable}

\end{document}